\documentclass[journal, draftclsnofoot, onecolumn, 12pt]{IEEEtran}
\usepackage{graphicx}
\usepackage{cite}
\usepackage{amssymb,amsmath}
\usepackage{algorithm}
\usepackage{algpseudocode}
\usepackage{epsfig}
\usepackage{epstopdf}
\usepackage{color}
\usepackage{dsfont}
\DeclareMathAlphabet{\mathpzc}{OT1}{pzc}{m}{it}

\begin{document}
\title{ Underlay Spectrum Sharing Techniques with In-band Full-Duplex Systems using Improper~Gaussian~Signaling}
\author{
\IEEEauthorblockN{\large  Mohamed Gaafar, Osama Amin, Walid Abediseid and Mohamed-Slim Alouini}

\thanks{The authors are with Computer, Electrical, and Mathematical Sciences and Engineering (CEMSE) Divison, King Abdullah University of Science and Technology (KAUST), Thuwal, Makkah Province, Saudi Arabia. Email: \{mohamed.gaafar, osama.amin, walid.abediseid , slim.alouini\}@kaust.edu.sa }}

\maketitle




\begin{abstract}
Sharing the spectrum with in-band full-duplex (FD) primary users (PU) is a challenging and interesting problem in the underlay cognitive radio (CR) systems. The self-interference introduced at the primary network may dramatically impede the secondary user (SU) opportunity to access the spectrum. To tackle this problem, we use the so-called \textit{improper Gaussian signaling}. Particularly, we assume a system with a SU pair working in a half-duplex mode that uses improper Gaussian signaling while the FD PU  pair implements the regular proper Gaussian signaling. First, we derive a closed form expression and an upper bound for the SU and PU outage probabilities, respectively. Second, we optimize the SU signal parameters to minimize its outage probability while maintaining the required PU quality-of-service based on the average channel state information (CSI). Moreover, we provide the conditions to reap merits from employing improper Gaussian signaling at the SU. Third, we design the SU signal parameters based on perfect knowledge of its direct link instantaneous CSI and investigate all benefits that can be achieved at both the SU and PU.  Finally, we provide some numerical results that demonstrate the advantages of using improper Gaussian signaling to access the spectrum of the FD PU.
\end{abstract}


\section{Introduction}

Cognitive radio (CR) is a promising spectrum sharing technology that mitigates the spectrum scarcity that resulted from the recent tremendous growth of wireless devices and services over the past decade \cite{mitola1999cognitive}. As many licensed primary users (PU) block the available spectrum, CR systems exploit  the same spectrum resources and allow secondary users (SU) to coexist with PU without degrading the PU quality-of-service (QoS) \cite{zhao2007survey}. CR can operate in three strategies called, interweave, overlay and underlay spectrum sharing. The CR with interweave strategy accesses the PU spectrum when it is not used, i.e., the PU is idle. On the other hand, CR with both overlay and underlay strategies coexists with the PU. In the overlay strategy, the SU uses part of its power to assist the PU transmission in order to compensate its interference impact on the PU. The scenario is more challenging for the underlay strategy, where the SU has to control its power to be within an acceptable level at the PU receiver side. CR systems can be incorporated with other communication techniques, such as full-duplex (FD) and cooperative communications, to improve the spectrum utilization of the communication networks and the SU performance.

FD is a spectral efficient paradigm that allows the communication nodes to  transmit and receive simultaneously at the same frequency. FD has recently attracted wide attention especially after the progress in self-interference cancellation, which gives a great promise in practical realization \cite{sabharwal2014band}. In underlay CR systems, FD is used to compensate the spectral efficiency loss of cooperative systems that is used to increase the SU coverage \cite{kim2012optimal, zhongTOAPPEARperformance}. In addition to that, FD is used to achieve simultaneous sensing and data transmission for SU, or possibly receive data from the other SU node during the transmission according to the channel conditions \cite{afifi2015incorporating}. The existing research on CR systems avoids sharing the spectrum with in-band FD PU due to the increased interference limitations at the PU side, which can impede the operation of underlay CR systems. Thus, investigating communication systems that can relieve the interference signature on the PU while improving the SU performance become imperative.

Improper Gaussian signaling has proven its superiority over proper Gaussian signaling to improve the achievable  rate in interference-limited networks  \cite{cadambe2010interference, ho2012improper, zeng2013transmit, zeng2013optimized, kurniawan2015improper}. In CR systems,  improper Gaussian signaling is employed in \cite{lameiro2015benefits, amin2015outage}, where the PU is assumed to work in half-duplex mode with proper Gaussian signaling. On the other hand, the SU is assumed to use improper Gaussian signaling and have perfect instantaneous channel state information (CSI) of all PU and SU communication links in \cite{lameiro2015benefits} and average CSI in \cite{amin2015outage}. Improper Gaussian signaling showed better performance than proper Gaussian signaling when the PU is not fully loaded. In \cite{gaafar2015spectrum}, we considered a challenging spectrum sharing scenario, where the PU is assumed to work using in-band full-duplex mode and we aim to improve the SU instantaneous rate performance while satisfying PU minimum fixed rate requirement. The main objective of the study in \cite{gaafar2015spectrum} is to investigate possible benefits for the improper Gaussian signaling scheme compared with the proper one, thus we assumed the availability of all CSI links at the SU side.

In this paper\footnote{This paper extends our work in \cite{gaafar2015sharing}, in which we provide more details and insights for the outage performance analysis and the average CSI based design. We also introduce a design based on the instantaneous SU direct link CSI and investigate its benefits for both SU and~PU.  Finally, more numerical results are presented.}, we assume also the spectrum sharing problems with FD licensed users, i.e., PU, and inspect the possibility of inserting the SU into operation without deteriorating the PU QoS but based on average or partial CSI. In this regards, we adopt the improper Gaussian signaling for the SU to create a room for spectrum sharing systems and measure the QoS by the outage probability for a target rate. To the best of our knowledge, this is the first work that investigates sharing the licensed spectrum of in-band FD PU based on average or partial CSI. The main contributions of this paper can
be summarized as follows:
 \begin{itemize}
 \item We derive a closed form expression for the outage probability of the SU, which employs improper Gaussian signaling and is subjected to interference from the FD PU, which employs proper Gaussian signaling.
 \item For the PU, which is subjected to SU and residual self-interference, we derive a tight upper bound for the outage probability in terms of the SU signal parameters represented in the SU power and the circularity coefficient. 
 \item We design the SU signal parameters to improve its performance based on average CSI while maintaining  acceptable PU QoS. In this context, the system performance is measured in terms of the outage probability. Moreover, we derive the conditions that can provide improvement via the usage of improper Gaussian signaling at the SU. 
  \item Then, we study the benefits of the partial instantaneous CSI at the SU terminals on the design problem. Specifically, we assume a practical scenario, in which perfect knowledge of SU direct link and average CSI of other links are available at the SU terminals. Then, we analyze the performance gain in  power saving, outage probability for both SU and PU, and SU average energy efficiency (EE).
 \item Finally, with the aid of numerical results, we first examine the accuracy of the PU derived outage probability bound. Then investigate the benefits that can be reaped by employing improper Gaussian signaling for the SU versus different system parameters. And finally, we explore the availability of the SU direct link benefits on power saving, SU and PU outage probability and SU average EE.
 \end{itemize}

The rest of the paper is organized as follows: In Section II, we introduce the system model of the spectrum sharing system with FD system. In Section III, we derive the outage probability expressions for both SU and PU. In Section IV, we design the SU signal parameters to minimize the SU outage probability subject to a given QoS for the PU based on average CSI. In Section V, we study the system performance with perfect CSI knowledge for the SU direct link and look for its advantages on both SU and PU. In Section VI, we present a comprehensive  simulation study to illustrate the  performance of the proposed improper Gaussian signaling techniques. Finally, Section VI concludes the paper.

\textbf{\textit{Notation:}} Throughout the rest of the paper, we use $ \left| {{\rm{ }}{\rm{. }}} \right| $ to denote the absolute value operation  and $\left[x \right]^{+} = \max(x,0)$. $\mathds{1}_{\left[ x, \infty \right)}\left( z \right)$ is the indicator function where $\mathds{1}_{\left[ x, \infty \right)}\left( z \right):=1$ if $z \geq x$ and $\mathds{1}_{\left[ x, \infty \right)}\left( z \right):=0$ for all $z < x$. $\Pr\{A\}$ denotes the probability of occurrence of an event A. The operator $\mathbb{E}\{.\}$ is used to denote the statistical expectation and the mean of a random variable (RV) $X$ is defined as $\bar{X} = \mathbb{E}\{ X \}$. 
\section{System Model}
\begin{figure}[!t]
\centering
\includegraphics[width=3in]{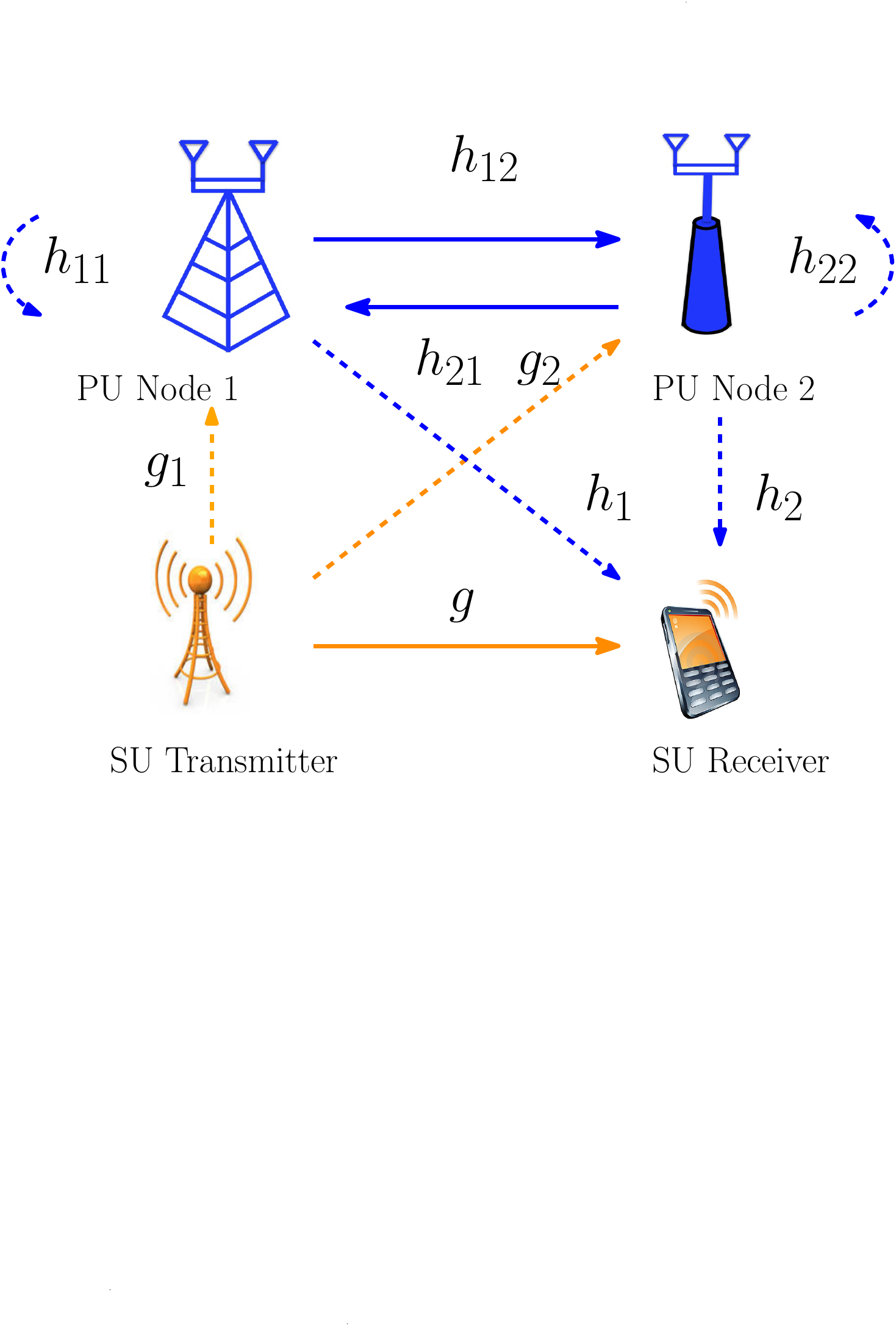}
\caption{System model, where the solid and the dash lines are for the desired and the interference signals, respectively.}
\label{fig1}
\end{figure}

Consider an underlay CR system, where a half-duplex SU pair coexists with an in-band full-duplex PU pair as depicted in Fig. \ref{fig1}. In this scenario, we have three simultaneous different transmissions occur in the same network over the same frequency. Before proceeding in describing the system components, we define the following terms:

\textit{Definition 1}: The variance and pseudo-variance of a zero mean complex random variable $x$ are defined as $\sigma _x^2 = {\mathbb{E}} {{{\{\left| x \right|}^2\}}} $ and $\hat{\sigma} _x^2 = {\mathbb{E}} {{\{{ x }^2\}}} $ respectively \cite{Neeser1993proper}.

\textit{Definition 2}: The proper signal has a zero  $\hat{\sigma} _x^2 $, while improper signal has a non-zero $\hat{\sigma} _x^2 $.  

\textit{Definition 3}: The circularity coefficient $\mathcal{C}_x$ measures the degree of impropriety of signal $x$ and is defined as $\mathcal{C}_x = \left|\hat{\sigma} _x^2 \right|/\sigma _x^2$, where $0 \le {\mathcal{C}_x} \le 1$. $\mathcal{C}_x=0$ denotes \textit{proper} signal and $\mathcal{C}_x=1$ denotes \textit{maximally improper} signal. 

The PU nodes are assumed to use zero-mean proper Gaussian signals $x_i, i \in {\{1,2\}}$ with a unit variance, while the SU transmitter uses zero-mean improper Gaussian signals $x_\mathrm{s}$ with a unit variance and a circularity coefficient $\mathcal{C}_x$. The received signal at the  PU node $j$, where $ j \in {\{1,2\}}$, $i \neq j$, is expressed~as
\begin{equation} \label{pu_sig}
{y_j} = \sqrt {{p_i}} {h_{ij}}{x_i} + \sqrt {{p_j}} {h_{jj}}{x_j} + \sqrt {{p_{\mathrm{s}}}} {g_{j}}{x_{\mathrm{s}}} + {n_j},  
\end{equation}
where $p_i$ is the transmitted power of the PU node $i$, ${p_{\mathrm{s}}}$  is the SU transmitted power, $n_j$ is the noise at the  PU node $j$ receiver, $h_{ij}$ denotes the communication channel between PU node $i$  and PU node $j$, $g_{j}$ represents the interference channel between the SU transmitter and PU node $j$ and $h_{jj}$ represents the residual self interference (RSI) channel of node $j$ after undergoing analog and digital cancellation techniques. We assume that the RSI is modeled as a zero mean complex Gaussian random variable as in \cite{kim2012optimal, day2012full}.  As for the PU direct links, we could assume channel reciprocity, i.e., $h_{ij}=h_{ji}$, however it might not be the case when the two PU nodes use different spatial antennas locations or the receivers' front end and transmitters' back end are not perfectly matched \cite{biglieri2007mimo}. Hence, we adopt the general assumption of different forward and reverse PU links. 

In the same time, the SU operates in half-duplex mode and the received signal is expressed~as
\begin{equation} \label{su_sig}
{y_{\rm{s}}} = \sqrt {{p_{\rm{s}}}} g{x_{\rm{s}}} + \sum\limits_{i = 1}^2 {\sqrt {{p_i}} {h_i}{x_i}}  + {n_{\rm{s}}},
\end{equation}
where $n_\mathrm{s}$ is the noise at the SU receiver, $h_{i}$ is the interference channel of PU node $i$ on the SU receiver, $g$ denotes the direct channel between the SU transmitter and receiver. The SU transmitter power has to be adjusted in order not to affect the required QoS of the PU. The channels in the described system are modeled as  slow Rayleigh flat fading channels  and the noise at the receivers end is modeled as a white, zero-mean, circularly
symmetric, complex Gaussian with variance~$\sigma^2$. The transmitted signals, channel coefficients and noise components are assumed to be independent of each other except $h_{ij}$ and $h_{ji}$, which may be dependent.

As a result of using improper Gaussian signaling while treating the interference as
Gaussian noise at all receivers, the achievable rate for the PU node $i$ is given by \cite{zeng2013transmit, lameiro2015benefits},
\begin{align}\label{pu_rate}
{R_{{{\mathrm{p}}_i}}}\left( {{p_{\mathrm{s}}},{\mathcal C_x}} \right) = {\log _2}\left( {1 + \frac{{{p_i}{\gamma _{{{\mathrm{p}}_i}}}}}{{{p_j}{\upsilon _{{{\mathrm{p}}_j}}} + {p_{\mathrm{s}}}{{\mathcal I}_{{{\mathrm{s}}_j}}} + 1}}} \right) + \frac{1}{2}{\log _2}\left( {\frac{{1 - {\mathcal C}_{{{{y}}_i}}^2}}{{1 - {\mathcal C}_{{{\mathcal{I}}_i}}^2}}} \right),
\end{align} 
where ${\gamma _{{{\mathrm{p}}_{_i}}}} = {{{{\left| {{h_{ij}}} \right|}^2}}} / {{{\sigma ^2}}}$ is the channel-to-noise ratio (CNR) of the PU channel from PU node $i$ to PU node $j$,  ${{\mathcal I}_{{{\mathrm{s}}_i}}} =  {{{{\left| {{g_{{i}}}} \right|}^2}}}/{{{\sigma ^2}}}$ is the interference CNR of SU to PU node $i$ and ${\upsilon _{{{\mathrm{p}}_i}}} = {{{{\left| {{h_{ii}}} \right|}^2}}}/{{{\sigma ^2}}}$ is the RSI-CNR of PU node $i$, ${\mathcal C}{_{{{y}_i}}}$ and ${{\mathcal C}_{{{\mathcal{I}}_i}}}$ are the circularity coefficients of the received and interference-plus-noise signals at PU node $i$, respectively, which are given by
\begin{align}\label{cir_coeff}
{\cal C}{_{{y_i}}} = \frac{{{p_{\rm{s}}}{{\cal I}_{{{\rm{s}}_j}}}{{\cal C}_x}}}{{{p_i}{\gamma _{{{{\rm{p}}}_i}}} + {p_j}{\upsilon _{{{{\rm{p}}}_j}}} + {p_{\rm{s}}}{{\cal I}_{{{\rm{s}}_j}}} + 1}}, \hspace{0.2cm} {{\cal C}_{{{\cal I}_i}}} = \frac{{{p_{\rm{s}}}{{\cal I}_{{{\rm{s}}_j}}}{{\cal C}_x}}}{{{p_j}{\upsilon _{{{{\rm{p}}}_j}}} + {p_{\rm{s}}}{{\cal I}_{{{\rm{s}}_j}}} + 1}}.
\end{align}

After some manipulations, ${R_{{{\mathrm{p}}_i}}}\left( {{p_{\mathrm{s}}},{\mathcal{C}_x}} \right)$ can be simplified as 
\begin{align}\label{pu_rate_simp}
 {R_{{{\mathrm{p}}_i}}}\left( {{p_{\mathrm{s}}},{\mathcal{C}_x}} \right) = \frac{1}{2}  {\log _2}\left( {\frac{{{{\left( {{p_i}{\gamma _{{{\mathrm{p}}_i}}} + {p_j}{\upsilon _{{{\mathrm{p}}_j}}} + {p_{\mathrm{s}}}{{\mathcal I}_{{{\mathrm{s}}_j}}} + 1} \right)}^2} - {{ {{p_{\mathrm{s}}^2}{{\mathcal I}_{{{\mathrm{s}}_j}}^2}{\mathcal{C}_x^2}} }}}}{{{{\left( {{p_j}{\upsilon _{{{\mathrm{p}}_j}}} + {p_{\mathrm{s}}}{{\mathcal I}_{{{\mathrm{s}}_j}}} + 1} \right)}^2} - {{ {{p_{\mathrm{s}}^2}{{\mathcal I}_{{{\mathrm{s}}_j}}^2}{\mathcal{C}_x^2}} }}}}} \right).
\end{align}

%

Similarly, the SU achievable rate can be expressed as
\begin{align}\label{su_rate}
& {R_{\mathrm{s}}}\left( {{p_{\mathrm{s}}},{\mathcal{C}_x}} \right) = \frac{1}{2}{\log _2}\left( {\frac{{p_{\mathrm{s}}^2{\gamma _{\mathrm{s}}}^2\left( {1 - {\mathcal{C}_x^2}} \right)}}{{{{\left( {\sum\nolimits_{i = 1}^2 {{p_i}{{\cal I}_{{{{\rm{p}}}_i}}}}  + 1} \right)}^2}}} + \frac{{2{p_{\mathrm{s}}}{\gamma _{\mathrm{s}}}}}{{\sum\nolimits_{i = 1}^2 {{p_i}{{\cal I}_{{{{\rm{p}}}_i}}}}  + 1}} + 1} \right),
\end{align}
where ${\gamma _{\mathrm{s}}} = {{{{\left| {{g}} \right|}^2}}}/{{{\sigma ^2}}}$ is the SU direct CNR between the SU transmitter and receiver and ${{\mathcal I}_{{{\mathrm{p}}_i}}} = {{{{\left| {{h_{i}}} \right|}^2}}}/{{{\sigma ^2}}}$ is the PU node $i$ interference CNR to the SU.

The direct, interference and RSI CNR ${\gamma _{{{\mathrm{p}}_{_i}}}}, \, { \gamma _{\mathrm{s}}}, \;{ {\mathcal I}_{{{\mathrm{p}}_i}}}, \; { {\mathcal I}_{{{\mathrm{s}}_i}}}, \, { \upsilon _{{{\mathrm{p}}_i}}}$  are  exponentially distributed random variables with mean values ${\bar \gamma _{{{\mathrm{p}}_{_i}}}}, \, {\bar \gamma _{\mathrm{s}}}, \, {\bar {\mathcal I}_{{{\mathrm{p}}_i}}}, \, {\bar {\mathcal I}_{{{\mathrm{s}}_i}}}, \, {\bar \upsilon _{{{\mathrm{p}}_i}}}$, respectively.

One can notice from  (\ref{pu_rate_simp}) and (\ref{su_rate}) that if $\mathcal{C}_x=0$, we obtain the well known achievable rates expressions of proper Gaussian signaling.  
Moreover, if $\mathcal{C}_x$ increases, the SU rate decreases while the PU rate increases  allowing the SU to increase its transmitted power. This merit should be considered carefully to satisfy the PU QoS requirements and meet the maximum SU power budget.
\section{Outage probability analysis}\label{out_prob_der}

Throughout this work, we use the outage probability to measure the performance for both the PU and SU assuming that both users use  fixed rates according to their transmission requirements. For this purpose, we derive expressions of the outage probability for both the PU and SU in the following subsections.
\subsection{Secondary User Outage Probability}
Assume the SU target rate is ${{R_{0,{\mathrm{s}}}}}$, then its outage probability is defined as
\begin{align}\label{su_out}
{P_{{\mathrm{out,s}}}}\left( {{p_{\mathrm{s}}},{{\mathcal C}_x}} \right) = \Pr \left\{ {{R_{\mathrm{s}}}\left( {{p_{\mathrm{s}}},{{\mathcal C}_x}} \right) < {R_{0,{\mathrm{s}}}}} \right\}.
\end{align}
By substituting (\ref{su_rate}) in (\ref{su_out}), we obtain
\begin{align}\label{su_out_ineq} 
& {P_{{\mathrm{out,s}}}}\left( {{p_{\mathrm{s}}},{{\mathcal C}_x}} \right) =  \Pr \Bigg\{ \frac{{p_{\mathrm{s}}^2{\gamma _{\mathrm{s}}^2}\left( {1 - {\mathcal{C}_x^2}} \right)}}{{{{\left( {\sum\nolimits_{i = 1}^2 {{p_{_i}}{{\cal I}_{{{{\rm{p}}}_i}}}}  + 1} \right)}^2}}} + \frac{{2{p_{\mathrm{s}}}{\gamma _{\mathrm{s}}}}}{\sum\nolimits_{i = 1}^2 {{p_{_i}}{{\cal I}_{{{{\rm{p}}}_i}}}}  + 1} - {\Gamma _{\mathrm{s}}} < 0 \Bigg\},
\end{align}
where ${\Gamma _{\mathrm{s}}} = {2^{2{R_{0,{\mathrm{s}}}}}} - 1$. One can show that the conditional SU outage probability (conditioned on ${{\mathcal I}_{{{\mathrm{p}}_1}}}$and ${{\mathcal I}_{{{\mathrm{p}}_2}}}$) is given by
\begin{align}\label{su_out_int}
{P_{{\mathrm{out,s}}}}\left( {{p_{\mathrm{s}}},{{\mathcal C}_x}\left| {{{\mathcal I}_{{{\mathrm{p}}_1}}},{{\mathcal I}_{{{\mathrm{p}}_2}}}} \right.} \right) =  1 -  \exp \left( { - \frac{{\gamma _{{{\mathrm{s}}}}^\circ}}{{{{\bar \gamma }_{\mathrm{s}}}}}} \right),
\end{align}
where ${\gamma _{{{\mathrm{s}}}}^\circ}$ is the non-negative zero obtained by solving the quadratic inequality in (\ref{su_out_ineq}) with respect to ${\gamma _{\mathrm{s}}}$, which is found to be
\begin{equation}
\gamma _{\rm{s}}^\circ= \frac{{\sum\nolimits_{i = 1}^2 {{p_i}{{\cal I}_{{{\rm{p}}_i}}} + 1} }}{{\left( {1 - {\cal C}_x^2} \right)}}\Psi_{\mathrm{s}} \left( {{p_{\rm{s}}},{{\cal C}_x}} \right),
\end{equation}
where $\Psi _{\mathrm{s}}\left( {p_{\mathrm{s}}},{{\mathcal C}_x} \right) = \left(\sqrt {1 + {\Gamma _{\mathrm{s}}}\left( {1 - {\mathcal{C}_x^2}} \right)}  - 1\right) / {{p_{\mathrm{s}}}}$. 
By averaging over the exponential statistics of ${{\mathcal I}_{{{\mathrm{p}}_i}}}$, we get
\begin{equation} \label{p_out_s}
{P_{{\rm{out}},{\rm{s}}}}\left( {{p_{\rm{s}}},{{\cal C}_x}} \right) = 1 - \frac{{\exp \left( { - \frac{{{\Psi _{\rm{s}}}\left( {{p_{\rm{s}}},{{\cal C}_x}} \right)}}{{{{\bar \gamma }_{\rm{s}}}\left( {1 - {\cal C}_x^2} \right)}}} \right)}}{{\prod\limits_{j = 1}^2 {\left( {{p_j}{{\overline {\cal I} }_{{{{\rm{p}}}_j}}}\frac{{{\Psi _{\rm{s}}}\left( {{p_{\rm{s}}},{{\cal C}_x}} \right)}}{{{{\bar \gamma }_{\rm{s}}}\left( {1 - {\cal C}_x^2} \right)}} + 1} \right)} }}.
\end{equation}
For the \textit{proper} signaling case, the above expression reduces~to
\begin{equation} \label{su_outc0}
{P_{{\rm{out}},{\rm{s}}}}\left( {{p_{\rm{s}}},0} \right) = 1 - \frac{{\exp \left( { - \frac{2^{R_{0,\rm{s}}}-1}{{{p_{\rm{s}}}{{\bar \gamma }_{\rm{s}}}}}} \right)}}{{\prod\limits_{j = 1}^2 {\left( {{p_j}{{\overline {\cal I} }_{{{{\rm{p}}}_j}}}\frac{2^{R_{0,\rm{s}}}-1}{{{p_{\rm{s}}}{{\bar \gamma }_{\rm{s}}}}} + 1} \right)} }}.
\end{equation}
while for \textit{maximally improper} case, the expression reduces to
\begin{equation}
{P_{{\rm{out}},{\rm{s}}}}\left( {{p_{\rm{s}}},1} \right) = \mathop {\lim }\limits_{{{\cal C}_x} \to 1} {P_{{\rm{out}},{\rm{s}}}}\left( {{p_{\rm{s}}},{{\cal C}_x}} \right) = 1 - \frac{{\exp \left( { - {\Gamma _{\rm{s}}}/2{p_{\rm{s}}}{{\bar \gamma }_{\rm{s}}}} \right)}}{{\prod\limits_{j = 1}^2 {\left( {\frac{{{p_j}{{\bar {\cal I} }_{{{{\rm{p}}}_j}}}{\Gamma _{\rm{s}}}}}{{2{p_{\rm{s}}}{{\bar \gamma }_{\rm{s}}}}} + 1} \right)} }}.\nonumber
\end{equation}
\subsection{Primary User Outage Probability}
The outage probability of PU node $i$ for a given target rate $R_{0,{{\mathrm{p}}_i}}$ is defined as
\begin{align}\label{pu_out}
{P_{{\mathrm{out}},{{\mathrm{p}}_i}}}\left( {{p_{\mathrm{s}}},\mathcal{C}_x} \right) = \Pr \left\{ {{R_{{{\mathrm{p}}_i}}}\left( {{p_{\mathrm{s}}},\mathcal{C}_x} \right) < {R_{0,{{\mathrm{p}}_i}}}} \right\}.
\end{align}
Substituting \eqref{pu_rate_simp} in \eqref{pu_out}, we obtain
\footnotesize{
\begin{align}\label{pu_out_ineq}
&{P_{{\rm{out,}}{{{\rm{p}}}_i}}}\left( {{p_{\rm{s}}},{{\cal C}_x}} \right) = \Pr \Bigg\{ \gamma _{{{{\rm{p}}}_i}}^2 + \frac{2}{{{p_{_i}}}}\left( {{p_{_j}}{\upsilon _{{{{\rm{p}}}_j}}} + {p_{\rm{s}}}{{\cal I}_{{{\rm{s}}_j}}} + 1} \right){\gamma _{{{{\rm{p}}}_i}}}  -\frac{{{\Gamma _{{{\rm{p}}}_i}}}}{{p_i^2}}\left( {{{\left( {{p_{_j}}{\upsilon _{{{{\rm{p}}}_j}}} + {p_{\rm{s}}}{{\cal I}_{{{\rm{s}}_j}}} + 1} \right)}^2} - p_{\rm{s}}^2{\cal I}_{{{\rm{s}}_j}}^2{\cal C}_x^2} \right) \le 0 \Bigg\},
\end{align}}
\normalsize
where ${\Gamma _{{{\rm{p}}}_i}} = {2^{2{R_{0,{{\rm{p}}_i}}}}} - 1$. Similar to the above subsection, the outage probability of PU node $i$ (conditioned on ${{\mathcal I}_{{{\mathrm{s}}_j}}}$ and ${\upsilon _{{{\mathrm{p}}_j}}}$) is given by
\begin{align}\label{pu_out_conditioned}
{P_{{\rm{out,}}{{{\rm{p}}}_i}}}\left( {{p_{\rm{s}}},{{\cal C}_x}\left| {{{\cal I}_{{{\rm{s}}_j}}},{\upsilon _{{{{\rm{p}}}_j}}}} \right.} \right) = 1 - \exp \left( { - \frac{{{{\gamma }_{{{\rm{p}}}_i}^\circ}}}{{{{\bar \gamma }_{{{\rm{p}}}_i}}}}} \right),
\end{align}
where ${{{\gamma }_{{{\rm{p}}}_i}^\circ}}$ represents the non-negative zero obtained by solving the quadratic inequality in (\ref{pu_out_ineq}) with respect to ${{ \gamma }_{{{\rm{p}}}_i}}$, and is found to be
\begin{align}\label{pu_out_int}
\gamma _{{{\rm{p}}}_i}^\circ  =  {\left( {{p_{_j}}{\upsilon _{{{{\rm{p}}}_j}}} + {p_{\rm{s}}}{{\cal I}_{{{\rm{s}}_j}}} + 1} \right)} {\Psi _{{{\mathrm{p}}_i}}}\left( {\frac{{{p_{\mathrm{s}}}{{\mathcal I}_{{{\mathrm{s}}_j}}}{{\mathcal C}_x}}}{{ {{p_{_j}}{\upsilon _{{{\mathrm{p}}_j}}} + {p_{\mathrm{s}}}{{\mathcal I}_{{{\mathrm{s}}_j}}} + 1} }}} \right),
\end{align}
where ${\Psi _{{{\mathrm{p}}_i}}}\left( x \right) = \left( {\sqrt {1 + {\Gamma _{{{\rm{p}}}_i}}\left( {1 - {x^2}} \right)}  - 1} \right)/ {{p_{_i}}} $.  
By averaging over the statistics of ${{\mathcal I}_{{{\mathrm{s}}_j}}}$ and ${\upsilon _{{{\mathrm{p}}_j}}}$ in \eqref{pu_out_conditioned}, we obtain
\begin{align} \label{pu_integral}
&{P_{{\mathrm{out,}}{{\mathrm{p}}_i}}}\left( {{p_{\mathrm{s}}},{{\mathcal C}_x}} \right) =  \mathbb{E}_{{{\mathcal I}_{{\mathrm{s}_j}}},{\upsilon _{{{\mathrm{p}}_j}}}}\left\{ {P_{{\mathrm{out,}}{{\mathrm{p}}_i}}}\left( {{p_{\mathrm{s}}},{{\mathcal C}_x}\left| {{{\mathcal I}_{{{\mathrm{s}}_j}}},{\upsilon _{{{\mathrm{p}}_j}}}} \right.} \right)  \right\} \nonumber \\
&=  1 - \int\limits_0^\infty  \int\limits_0^\infty  \frac{{\exp \left( { - \frac{x}{{{{\bar {\cal I}}_{{{\rm{s}}_j}}}}}} \right)\exp \left( { - \frac{y}{{{{\bar \upsilon }_{{{{\rm{p}}}_j}}}}}} \right)}}{{{{\bar {\cal I}}_{{{\rm{s}}_j}}}{{\bar \upsilon }_{{{{\rm{p}}}_j}}}}}  
 \exp \left( { - \frac {{p_{_j}}y + {p_{\rm{s}}}x + 1}{{{\bar \gamma }_{{{\rm{p}}}_i}}} {\Psi _{{{{\rm{p}}}_i}}}\left( {\frac{{x{p_{\rm{s}}}{\cal C}_x}}{{\left( {{p_{_j}}y + {p_{\rm{s}}}x + 1} \right)}}} \right)} \right)  dxdy.
\end{align}
Unfortunately, there is no closed form expression for the aforementioned integral except at $\mathcal{C}_x=0$, which yields
\begin{align}  \label{Pout_proper_PU}
{P_{{\rm{out,}}{{\rm{p}}_i}}} \left( {{p_{\rm{s}}},0} \right) = 1 -  \left( {\frac{{p_i^2\bar \gamma _{{{\rm{p}}_i}}^2\exp \left( { - \frac{{2^{{R_{0,{{\rm{p}}_i}}}}} - 1}{{{{p_i\bar \gamma }_{{{\rm{p}}_i}}}}}} \right)}}{{\left( {{p_{\rm{s}}}{{\bar {\cal I}}_{{{\rm{s}}_j}}}\left({2^{{R_{0,{{\rm{p}}_i}}}}} - 1\right)+ {{p_i\bar \gamma }_{{{\rm{p}}_i}}}} \right)\left( {{p_{_j}}{{\bar \upsilon }_{{{\rm{p}}_j}}}\left({2^{{R_{0,{{\rm{p}}_i}}}}} - 1\right) + {{p_i\bar \gamma }_{{{\rm{p}}_i}}}} \right)}}} \right).
\end{align}
For $\mathcal{C}_x\neq 0$, we resort to obtaining an upper bound for the outage probability. First, we study the convexity of the exponential term in \eqref{pu_out_conditioned} with respect to 
${ {\mathcal I}}_{{{\mathrm{s}}_j}}$ by defining $F\left( { {\mathcal I}}_{{{\mathrm{s}}_j}} \right) = \exp \left( {  G\left( { {\mathcal I}}_{{{\mathrm{s}}_j}} \right)} \right)$, where $ G\left( {  {\mathcal I}}_{{{\mathrm{s}}_j}} \right) = \left( {D{ {\mathcal I}}_{{{\mathrm{s}}_j}} + F} \right) - \sqrt {A{{  {\mathcal I}}_{{{\mathrm{s}}_j}}^2} + B{  {\mathcal I}}_{{{\mathrm{s}}_j}} + C} $ and $A = \frac{{p_{\rm{s}}^2\left( {1 + {\Gamma _{{\rm{p}_i}}}\left( {1 - {\cal C}_x^2} \right)} \right)}}{{p_i^2\bar \gamma _{{\rm{p}_i}}^2}}$, $B = \frac{{2{p_{\rm{s}}}\left( {1 + {\Gamma _{{\rm{p}_i}}}} \right)\left( {1 + {p_j}{\upsilon _{{{\rm{p}}_j}}}} \right)}}{{p_i^2\bar \gamma _{{\rm{p}_i}}^2}}$, $ C = \frac{{\left( {1 + {\Gamma _{{\rm{p}_i}}}} \right){{\left( {1 + {p_j}{\upsilon _{{{\rm{p}}_j}}}} \right)}^2}}}{{p_i^2\bar \gamma _{{\rm{p}_i}}^2}}$, $D = \frac{{{p_{\rm{s}}}}}{{{p_i}{{\bar \gamma }_{{\rm{p}_i}}}}}$ and $F = \frac{{1 + {p_j}{\upsilon _{{{\rm{p}}_j}}}}}{{{p_i}{{\bar \gamma }_{{\rm{p}_i}}}}}$ are positive constants. The second derivative of $G\left( {{\mathcal I}_{{{\mathrm{s}}_j}}} \right)$ is given by
\begin{align}
\frac{{{\partial ^2}G\left( {{\mathcal I}_{{{\mathrm{s}}_j}}} \right)}}{{\partial {{\mathcal I}_{{{\mathrm{s}}_j}}^2}}} = \frac{{{B^2}-4AC}}{{4{{\left( {C + {{\mathcal I}_{{{\mathrm{s}}_j}}}\left( {B + A{{\mathcal I}_{{{\mathrm{s}}_j}}}} \right)} \right)}^{3/2}}}}>0,
\end{align}
which proves the convexity of $G\left( {{\mathcal I}_{{{\mathrm{s}}_j}}} \right)$ and  $F\left( {{\mathcal I}_{{{\mathrm{s}}_j}}} \right)$ \cite{boyd2004convex}. As a result, we can employ the Jensen's inequality \cite{gradstein1980tables} over ${{\mathcal I}_{{{\mathrm{s}}_j}}}$ obtaining a conditional upper bound as  
\begin{align} \label{pu_cond_ub}
&P_{{\rm{out,}}{{{\rm{p}}}_i}}\left( {{p_{\rm{s}}},{{\cal C}_x}} | \upsilon _{{{{\rm{p}}}_j}} \right) \leq 1 -   \exp  \left( - \frac{{{p_{_j}}{\upsilon _{{{{\rm{p}}}_j}}} + {p_{\rm{s}}}{{\bar{\cal I}}_{{{\rm{s}}_j}}} + 1}}{{{{\bar \gamma }_{{{\rm{p}}}_i}}}} {\Psi _{{{{\rm{p}}}_i}}}\left( {\frac{{{p_{\rm{s}}}{{\bar{\cal I}}_{{{\rm{s}}_j}}}{{\cal C}_x}}}{{ {{p_{_j}}{\upsilon _{{{{\rm{p}}}_j}}} + {p_{\rm{s}}}{\bar{{\cal I}}_{{{\rm{s}}_j}}} + 1} }}} \right) \right).
\end{align}
Similarly, one can prove that the exponential term in  \eqref{pu_cond_ub} is convex in $\upsilon _{{{{\rm{p}}}_j}}$ and obtain the following upper bound of the PU outage probability
\begin{align} \label{pu_outage_ub}
&P_{{\rm{out,}}{{{\rm{p}}}_i}}\left( {{p_{\rm{s}}},{{\cal C}_x}} \right) \leq 1 -   \left( - \frac{{{p_{_j}}{\bar \upsilon _{{{{\rm{p}}}_j}}} + {p_{\rm{s}}}{{\bar{\cal I}}_{{{\rm{s}}_j}}} + 1}}{{{{\bar \gamma }_{{{\rm{p}}_i}}}}} {\Psi _{{{{\rm{p}}}_i}}}\left( {\frac{{{p_{\rm{s}}}{{\bar{\cal I}}_{{{\rm{s}}_j}}}{{\cal C}_x}}}{{ {{p_{_j}}{\bar \upsilon _{{{{\rm{p}}}_j}}} + {p_{\rm{s}}}{\bar{{\cal I}}_{{{\rm{s}}_j}}} + 1} }}} \right) \right)  \triangleq P_{{\mathrm{out,}}{{\mathrm{p}}_i}}^{\mathrm{UB}}\left( {{p_{\mathrm{s}}},{{\mathcal C}_x}} \right). 
\end{align}

\section{Secondary User Transmitted Signal Design}
In this section, we optimize the SU signal parameters to minimize the SU outage probability while maintaining a predetermined PU outage probability for each PU link. 

First, we state a unified PU design criterion in order to be satisfied by the SU during the operation with either proper or improper Gaussian signaling. Assume the PU nodes transmit with a fixed rate ${R_{0,{{{\rm{p}}}_i}}}$ and a target maximum outage probability threshold of  ${{\cal O}_{{{{\rm{p}}}_i}}}$ while allowing an interference power margin ${{\cal P}_{{\rm{int,}}{{{\rm{p}}}_i}}}$. Although the licensed band is dedicated locally only to the PU, the interference protection does not only mitigate the RSI but it can  also mitigate other interference that can come from other sources such as frequency reuse users in cellular networks \cite{holma2010wcdma}. The interference margin is used to ensure certain link budget that is suitable for the transmission QoS. The PU is assumed to use the maximum power budget $p_i$ to guarantee achieving its required QoS. Thus, the PU outage probability, from its perspective, is expressed~as  
\begin{align} \label{pu_out_margin}
{{\cal O}_{{{{\rm{p}}}_i}}} &= \Pr \left\{ {{{\log }_2}\left( {1 + \frac{{{p_i}{{\left| {{h_{ij}}} \right|}^2}}}{{{\sigma ^2} + {{\cal P}_{{\rm{int,}}{{{\rm{p}}}_i}}}}}} \right) < {R_{0,{{{\rm{p}}}_i}}}} \right\}  = 1 - \exp \Bigg( { -  \frac{{1 + {{\cal I}_{\max ,{{{\rm{p}}}_i}}}}}{{{{p_i\bar \gamma }_{{{\rm{p}}}_i}}}}\left({2^{{R_{0,{{\rm{p}}_i}}}}} - 1\right) } \Bigg),
\end{align}
where ${{\cal I}_{\max ,{{{\rm{p}}}_i}}}={{\cal P}_{{\rm{int,}}{{{\rm{p}}}_i}}}/{\sigma ^2}$ is the maximum allowable margin interference-to-noise ratio (INR) at the receiver of PU node $i$. By considering a maximum PU outage probability threshold ${{\cal O}_{{{{\rm{p}}}_i}}}$, ${{\cal I}_{\max ,{{{\rm{p}}}_i}}}$ can be found from \eqref{pu_out_margin} as
\begin{align}
{{\cal I}_{\max ,{{{\rm{p}}}_i}}} = \left[\frac{{{\mu _i}}}{{\sqrt {1 + {\Gamma _{{{\rm{p}}}_i}}}  - 1}} - 1\right]^+,
\end{align}
where ${\mu _i} = {p_i}{{\bar \gamma }_{{{\rm{p}}}_i}}\log \left( {\frac{1}{{1 - {{\cal O}_{{p_i}}}}}} \right)$ and $\log\left(.\right)$ is the natural logarithm. In the following subsections, we design proper and improper Gaussian signals for the SU to improve its performance,  measured by the outage probability, considering a predetermined ${{\cal O}_{{{{\rm{p}}}_i}}}$ and other system parameters such as $p_i$ and ${R_{0,{{{\rm{p}}}_i}}}$.
\subsection{Proper Gaussian Signaling Design}
In the case of proper Gaussian signaling, the SU allocates its power in order to minimize its outage probability subject to its own power budget ${p_{{\rm{s,max}}}}$ and PU QoS by solving the following  optimization problem,
\begin{align} \label{opt_prob_proper}
 & \mathop {\min }\limits_{{p_{\rm{s}}}} \quad {P_\mathrm{out,s}}\left( {{p_{\rm{s}}}, 0} \right) \hspace{0.25cm} \nonumber \\
& \hspace{0.1cm}\mathrm{s. \; t.} \quad  {P_{\mathrm{out},{{\mathrm{p}}_i}}}\left( {{p_{\mathrm{s}}},0} \right) \le {{\cal O}_{{p_i}}}, \nonumber \\
&\quad \quad \quad 0 < {p_{\mathrm{s}}} \le {p_{{\mathrm{s,max}}}}.
\end{align}
The predetermined PU outage probability  constraints in \eqref{opt_prob_proper} reduce to ${p_{\rm{s}}}\leq{p^{\left(i\right)}_{\rm{s}}}$, where ${p^{\left(i\right)}_{\rm{s}}}$ is the maximum allowable power that satisfies the PU required  outage probability threshold, which is expressed as 
\begin{align} \label{proper_psi}
p_{\rm{s}}^{\left( i \right)} = {\left[ {\frac{{\exp \left( { - \frac{{2^{{R_{0,{{\rm{p}}_i}}}}} - 1}{{{{p_i\bar \gamma }_{{{{\rm{p}}}_i}}}}}} \right) - \left( {1 - {\cal{O}_{{{{\rm{p}}}_i}}}} \right)\left( {{p_{_j}}{{\bar \upsilon }_{{{{\rm{p}}}_j}}}\frac{{{2^{{R_{0,{{\rm{p}}_i}}}}} - 1}}{{{{p_i\bar \gamma }_{{{{\rm{p}}}_i}}}}} + 1} \right)}}{{{{\bar {\cal I}}_{{{\rm{s}}_j}}}\left( {1 - {\cal{O}_{{{{\rm{p}}}_i}}}} \right)\frac{{{2^{{R_{0,{{\rm{p}}_i}}}}} - 1}}{{{{p_i\bar \gamma }_{{{{\rm{p}}}_i}}}}}\left( {{p_{_j}}{{\bar \upsilon }_{{{{\rm{p}}}_j}}}\frac{{{2^{{R_{0,{{\rm{p}}_i}}}}} - 1}}{{{{p_i\bar \gamma }_{{{{\rm{p}}}_i}}}}} + 1} \right)}}} \right]^ + }.
\end{align}
Thus, we can rewrite the optimization problem in \eqref{opt_prob_proper} as  
\begin{align} \label{opt_prob_proper_equivalent}
 & \mathop {\min }\limits_{{p_{\rm{s}}}} \quad {P_\mathrm{out,s}}\left( {{p_{\rm{s}}}, 0} \right) \hspace{0.25cm} \nonumber \\
& \hspace{0.1cm}\mathrm{s. \; t.} \quad  p_{\rm{s}} \le \min \left( {p_{\rm{s}}^{\left( 1 \right)},p_{\rm{s}}^{\left( 2 \right)},{p_{{\rm{s}},{\rm{max}}}}} \right).
\end{align}
One can prove that ${P_\mathrm{out,s}}\left( {{p_{\rm{s}}}, 0} \right)$ is monotonically decreasing in $p_\mathrm{s}$, thus the upper bound of the constraint achieves the optimal minimum SU outage probability and is expressed as 
\begin{align} \label{ps_proper_avg_case}
{p_{\rm{s}}} = \mathop {\min } \left( p_{\rm{s}}^{\left( 1 \right)}, p_{\rm{s}}^{\left( 2 \right)}, p_{\mathrm{s,max}} \right).
\end{align}
From (\ref{proper_psi}) and \eqref{ps_proper_avg_case}, the SU operates if ${\exp \left( { - \frac{{2^{{R_{0,{{\rm{p}}_i}}}}} - 1}{{{{p_i\bar \gamma }_{{{{\rm{p}}}_i}}}}}} \right) > \left( {1 - {\cal{O}_{{{{\rm{p}}}_i}}}} \right)\left( {{p_{_j}}{{\bar \upsilon }_{{{{\rm{p}}}_j}}}\frac{{{2^{{R_{0,{{\rm{p}}_i}}}}} - 1}}{{{{p_i\bar \gamma }_{{{{\rm{p}}}_i}}}}} + 1} \right)}$, which reduces to the following spectrum sharing condition for the maximum allowable margin INR as 
\begin{align}
{{\cal I}_{\max ,{{{\rm{p}}}_i}}} > \frac{{{{p_i\bar \gamma }_{{{{\rm{p}}}_i}}}}}{{2^{{R_{0,{{\rm{p}}_i}}}}} - 1}\log \left( {1 + {p_j}{{\bar \upsilon }_{{{{\rm{p}}}_j}}}\frac{{{2^{{R_{0,{{\rm{p}}_i}}}}} - 1}}{{{{p_i\bar \gamma }_{{{{\rm{p}}}_i}}}}}} \right).
\end{align}   
Otherwise, the SU should remain silent. Therefore, we develop Algorithm I to allocate the optimal power for the SU transmission.
\floatname{algorithm}{}
\begin{algorithm} \label{alg1}
\renewcommand{\thealgorithm}{}
\caption{\textbf{Algorithm I}}
\begin{algorithmic}[1]
\State \textbf{Input} $p_i$, ${\bar \gamma _{{{\mathrm{p}}_{_i}}}}$, ${\bar{ \mathcal I}_{{{\mathrm{s}}_i}}}$, ${\bar \upsilon _{{{\mathrm{p}}_i}}}$, ${R_{0,{{{\rm{p}}}_i}}}$, ${{\cal O}_{{p_i}}}$, $p_{\mathrm{s}}^{(0)} = {p_{{\rm{s,max}}}}$
\If{${{\cal I}_{\max ,{{{\rm{p}}}_i}}} > \frac{{{{p_i\bar \gamma }_{{{{\rm{p}}}_i}}}}}{{2^{{R_{0,{{\rm{p}}_i}}}}} - 1}\log \left( {1 + {p_j}{{\bar \upsilon }_{{{{\rm{p}}}_j}}}\frac{{{2^{{R_{0,{{\rm{p}}_i}}}}} - 1}}{{{{p_i\bar \gamma }_{{{{\rm{p}}}_i}}}}}} \right)$}
\State \textbf{Compute} $m = \mathop {\arg \min }\limits_{l \in \{0,1,2\}} \hspace{0.2cm}  p_{\mathrm{s}}^{( l  )} $ 
\State \textbf{Output} $p_{\rm{s}}^*=p_{\mathrm{s}}^{( m )}$
\Else {}
\State \textbf{Output} $p_{\rm{s}}^*=0$
\EndIf
\end{algorithmic}
\end{algorithm} 
  
\subsection{Improper Gaussian Signaling Design}
The improper Gaussian signal design aims to tune  $p_{\mathrm{s}}$ and $\mathcal{C}_x$ to minimize the SU outage while holding a required PU QoS based on the upper bound derived  in \eqref{pu_outage_ub}  achieving the worst case system design. To this end, we formulate the design optimization problem as,
\begin{align} \label{opt_prob}
&\mathop {\min }\limits_{{p_{\rm{s}}},{{\cal C}_x}} \quad {P_\mathrm{out,s}}\left( {{p_{\rm{s}}},{{\cal C}_x}} \right) \hspace{0.25cm} \nonumber \\
&\hspace{0.1cm}\mathrm{s. \; t. \;} \quad {P_{\mathrm{out},{{{\rm{p}}}_i}}^{\rm{UB}}}\left( {{p_{\rm{s}}},\;{\cal C}_x} \right) \le {{\cal O}_{{p_i}}}, \nonumber \\
& \quad \quad \quad \;0 < {p_{\rm{s}}} \leq {p_{{\rm{s,max}}}}, \nonumber \\
& \quad \quad \quad \;0 \leq {{\cal C}_x} \leq 1.
\end{align}
Unfortunately, this optimization problem turns to be non-linear and non-convex which makes it hard in general to find its optimal global solution. However, similar to what we did in the proper signaling design, we exploit some monotonicity properties of the objective function and the constraints that lead us to the optimal global solution of \eqref{opt_prob}. 

First, based on \eqref{pu_outage_ub}, we express the outage probability constraints in \eqref{opt_prob}  by the following equivalent quadratic inequality in terms of~${p_{\rm{s}}}$ as
\begin{align}\label{ps_improper_quadraric}
{\Gamma _{{{\rm{p}}}_i}}\left( {1 - \mathcal{C}_x^2} \right)\bar {\cal I} _{{{\rm{s}}_j}}^2 p_\mathrm{s}^2 + 2{\Lambda _i} {\bar {\cal I} _{{{\rm{s}}_j}}} p_\mathrm{s} - {\Upsilon _i} \le 0,
\end{align}
where ${\Upsilon _i} = \left( {\mu _i^2 + 2{\beta _j}{\mu _i} - {\Gamma _{{{\rm{p}}}_i}}\beta _j^2} \right)$, ${\Lambda _i} = \left({\beta _j}{\Gamma _{{{\rm{p}}}_i}} - {\mu _i}\right)$, ${\beta _j} = \left({p_j}{{\bar \upsilon }_{{{{\rm{p}}}_j}}} + 1\right)$. Based on \eqref{ps_improper_quadraric}, the outage probability constraints is equivalent to ${p_{\rm{s}}}\leq{p^{ (i )}_{\rm{s}}}\left({{\cal C}_x}\right)$, where ${p_{\rm{s}}^{(i)}}\left({{\cal C}_x}\right)$ is found by equating the left-hand-side of \eqref{ps_improper_quadraric} to zero and then compute the feasible root(s). One can show that if $\Upsilon _i \leq 0$, then, $ {\Lambda _i}>0$, which results in two negative roots. On the other hand, if $\Upsilon _i > 0$, then, there is exactly one positive root and one negative root. As a result, the feasible  power bound is given by  
\begin{align}\label{ps_improper_max}
p_\mathrm{s}^{\left( i \right)}\left({{\cal C}_x}\right) =  \left[\frac{{\sqrt {\Lambda _i^2 + {\Gamma _{{{\rm{p}}}_i}}\left( {1 - {\cal C}_x^2} \right){\Upsilon _i}}  - {\Lambda _i}}}{{{\Gamma _{{{\rm{p}}}_i}}{{\bar {\cal I}}_{{{\rm{s}}_j}}}\left( {1 - {\cal C}_x^2} \right)}}\right]^+.
\end{align}
Thus, the first three constraints in \eqref{opt_prob} can be equivalently rewritten as 
\begin{align}\label{ps_min}
{p_{\rm{s}}}\left({{\cal C}_x}\right) \le \min \left\{ {p_{\rm{s}}^{\left( 1 \right)}\left({{\cal C}_x}\right),\;p_{\rm{s}}^{\left( 2 \right)}}\left({{\cal C}_x}\right),\;{p_{{\rm{s,max}}}} \right\}. 
\end{align}
From \eqref{ps_improper_max} and \eqref{ps_min}, the SU is  allowed to transmit if $\Upsilon _i > 0$, which is always valid as long as ${{\cal I}_{\max ,{{{\rm{p}}}_i}}}>{p_j}{{\bar \upsilon }_{{{{\rm{p}}}_j}}}$. To obtain the distinct intersection points of the aforementioned piece-wise function  in $0 < {{\cal C}_x} < 1$, first, we can show that $p_{\rm{s}}^{\left( i \right)}$ is strictly increasing in ${\cal C}_x$\footnote{See Appendix A for the proof} over the interested interval. Hence, \eqref{ps_min} can be described with a maximum of four intervals (three breaking points) and a minimum of one interval (zero breaking points). The intersection point, $r^{\left( i \right)}$, between $p_{\rm{s}}^{\left( i \right)}$ and ${p_{{\rm{s,max}}}}$ is found from
\begin{align}\label{ps_i_ps_max_intersection}
r^{\left( i \right)} = \sqrt {1 + \frac{{2\left( {{p_{{\rm{s,max}}}}{{\bar {\cal I}}_{{{\rm{s}}_j}}}} \right){\Lambda _i} - {\Upsilon _i}}}{{{\Gamma _{{{\rm{p}}}_i}}{{\left( {{p_{{\rm{s,max}}}}{{\bar {\cal I}}_{{{\rm{s}}_j}}}} \right)}^2}}}},  
\end{align}   
which exists  if $p_s^{\left( i \right)}\left(0\right) < {p_{{\rm{s,max}}}}$ and $p_\mathrm{s}^{\left( i \right)}\left(1\right) > {p_{{\rm{s,max}}}}$.
Furthermore, the intersection between $p_{\rm{s}}^{\left( 1 \right)}$ and $p_{\rm{s}}^{\left( 2 \right)}$ in the interested interval, if they are not identical, is $r^{\left( 3 \right)} = \sqrt{(1 - \kappa)}$, where $\kappa$ is computed from
\begin{align} \label{ps_i_intersection}
\kappa = \frac{{4{\Gamma _{{\mathrm{p}}_1}}{\Gamma _{{\mathrm{p}}_2}}{{\bar {\cal I}}_{{{\rm{s}}_1}}}{{\bar {\cal I}}_{{{\rm{s}}_2}}}\left( {{\Gamma _{{\mathrm{p}}_1}}{\Lambda _2}{{\bar {\cal I}}_{{{\rm{s}}_2}}} - {\Gamma _{{\mathrm{p}}_2}}{\Lambda _1}{{\bar {\cal I}}_{{{\rm{s}}_1}}}} \right)\left( {{\Lambda _2}{\Upsilon _1}{{\bar {\cal I}}_{{{\rm{s}}_1}}} - {\Lambda _1}{\Upsilon _2}{{\bar {\cal I}}_{{{\rm{s}}_2}}}} \right)  }}{{{{\left( {{\Gamma _{{\mathrm{p}}_2}}{\Upsilon _1}\bar {\cal I}_{{{\rm{s}}_1}}^2 - {\Gamma _{{\mathrm{p}}_1}}{\Upsilon _2}\bar {\cal I}_{{{\rm{s}}_2}}^2} \right)}^2}}}.
\end{align} 
which exists if $p_s^{\left( i \right)}\left( 0 \right) < p_s^{\left( j \right)}\left( 0 \right)$ and $p_s^{\left( i  \right)}\left( 1 \right) > p_s^{\left( j \right)}\left( 1 \right)$. The identical case i.e., $p_{\rm{s}}^{\left( 1 \right)}=p_{\rm{s}}^{\left( 2 \right)}$ occurs if the links' received CNR at the receiver of PU node $i$ are equal. In such scenario, we have at most two intervals (one breaking point) that can be obtained from \eqref{ps_i_ps_max_intersection} if exists.    

Define the interval boundaries   points as ${\cal C}_ x^{(z)}$, where $z$ is an integer number in  $ [1,k+1]$, $k$ is the number of  distinct  intersection points, i.e. $k \in \{ 0,1,2,3 \}$, ${\cal C}_ x ^{(0)}=0$, ${\cal C}_x ^{(k+1)}=1$ and ${\cal C}_ x^{(1)} $, ${\cal C}_ x^{(2)} $ and ${\cal C}_ x^{(3)} $ are the ordered  distinct  intersection points (if exist).

Thereafter, we divide the optimization problem in  \eqref{opt_prob} into $(k+1)$ sub-problems, where  each sub-problem is defined in a specific range ${\cal C}_x ^{(z-1)} \le {{\cal C}_x} \le {\cal C}_x ^{(z )}$. We can show that ${P_\mathrm{out,s}}\left( p_{\mathrm{s}}, {{{\cal C}_x}} \right)$ is monotonically decreasing in $p_{\mathrm{s}}$ for a fixed ${{{\cal C}_x}}$. Hence, $p_\mathrm{s}$ is assigned the upper bound of \eqref{ps_min}  to minimize the outage probability. Then, we check the minimum of the three functions in \eqref{ps_min} in each interval and substitute the value of $p_{\rm{s}}$ in ${P_\mathrm{out,s}}\left( p_{\mathrm{s}}, {{{\cal C}_x}} \right)$ obtaining $k+1$ sub-problems, where the $z^{\mathrm{th}}$ sub-problem is written as
\begin{align} \label{opt_subprob}
\mathfrak{P}_{z}: & \mathop {\min }\limits_{{{\cal C}_x}}  \; {P_\mathrm{out,s}}\left( {{{\cal C}_x}} \right) \nonumber \\ & \mathrm{s. \; t. \; } \; {\cal C}_x^{(z-1)} < {{\cal C}_x} \le {\cal C}_ x ^{(z)}.
\end{align}
To solve the $\mathfrak{P}_{z}$ problem, we have two cases, either $p_{\rm{s}}={p_{{\rm{s,max}}}}$ or $p_{\rm{s}}=p_{\rm{s}}^{\left( m \right)}\left({{\cal C}_x}\right)$, where $m$ denotes the index of the minimum of the two functions $p_{\rm{s}}^{\left( i \right)}$ in the interested interval. \textit{Firstly}, if $p_{\rm{s}}= p_{\rm{s}}^{\left( m \right)}$ in \eqref{ps_min}, ${P_\mathrm{out,s}}\left( {{{\cal C}_x}} \right)$ reduces to
\begin{align}\label{p_out_s_Cx}
{P_\mathrm{out,s}}\left( {{{\cal C}_x}} \right) = 1 - \frac{{{\mathcal{Y}^2}{\mathcal{G}_m^2\left(\bar\gamma_{\rm{s}}\right)}}\exp \left( { - \frac{1}{{\mathcal{Y}\mathcal{G}_m\left(\bar\gamma_{\rm{s}}\right)}}} \right)}{ \prod\limits_{j = 1}^2 {\left( {{p_j}{{\bar {\cal I}}_{{{{\rm{p}}}_j}}} + \mathcal{Y}\mathcal{G}_m\left(\bar\gamma_{\rm{s}}\right)} \right)} },
\end{align}
where $\mathcal{Y}  = \sqrt {1 - {\cal C}_x^2} /\left( {\sqrt {1 + \left( {1 - {\cal C}_x^2} \right){\Gamma _{\rm{s}}}}  - 1} \right)$ and $\mathcal{G}_m\left(z\right)  = p_{\rm{s}}^{(m)}z\sqrt {1 - {\cal C}_x^2}$. Thus, ${P_\mathrm{out,s}}$ in \eqref{p_out_s_Cx} is monotonically decreasing in $\mathcal{C}_x$ if\footnote{See Appendix B for the proof.}
\begin{align} \label{improper_condition}
{R_{0,{\rm{s}}}} \leq {\log _2}\left( {\frac{{{\mu _m}\sqrt {1 + {\Gamma _{{\rm{p}_m}}}} }}{{{\beta _j}{\Gamma _{{\rm{p}_m}}} - {\mu _m}}}} \right), \qquad j \neq m.
\end{align} 
Therefore,  the optimal solution pair in this case is $ ( {p_{\rm{o}}^{(z)},{\cal C}_{\rm{o}}^{(z)}} )= (p_s^{\left( m \right)} ({\cal C}_ x ^{(z)} ), {\cal C}_x ^{(z)} )$.  Otherwise, it is monotonically increasing and hence,  $ ( {p_{\rm{o}}^{(z)},{\cal C}_{\rm{o}}^{(z)}} )= (p_s^{\left( m \right)} ({\cal C}_ x ^{(z-1)} ), {\cal C}_x ^{(z-1)} )$. \\ \textit{Remark :} Given that the maximum INR for PU node $i$ exceeds ${p_j}{{\bar \upsilon }_{{{{\rm{p}}}_j}}}$, the inequality in \eqref{improper_condition} gives the condition for improper Gaussian signals to be beneficial for the SU transmission.

\textit{Secondly}, if $p_{\rm{s}}={p_{{\rm{s,max}}}}$ in \eqref{ps_min}, then by substituting its value in the SU outage probability, we obtain ${P_\mathrm{out,s}}\left( {{{\cal C}_x}}\right)= {P_\mathrm{out,s}}\left( p_{\mathrm{s,max}}, {{{\cal C}_x}} \right)$ that is monotonically increasing in $\mathcal{C}_x$ and can be easily shown in a similar way to the proof in Appendix B. Hence, the optimal solution pair is $( {p_{\rm{o}}^{(z)},{\cal C}_{\rm{o}}^{(z)}} )= ({p_{{\rm{s,max}}}}, {\cal C}_ x^{(z-1)})$. At the end, we pick the global optimal pair $( {p_{\rm{o}}^{(z)},{\cal C}_{\rm{o}}^{(z)}} )$ that minimizes the objective function ${P_\mathrm{out,s}}\left( {{p_{\rm{s}}},{{\cal C}_x}}\right)$. Based on the aforementioned analysis, we develop Algorithms II to find the distinct intersection points and the optimal solution pairs in $z$ regions, then find the pair, $\left( {p_{\rm{s}}^*,{\cal C}_x^*} \right)$, with minimum SU outage probability from 
\begin{align}\label{opt_prob_global}
\left( {p_{\rm{s}}^*,{\cal C}_x^*} \right) = \mathop {\arg \min } \; {p_{{\rm{out}},{\rm{s}}}}\left( {p_{\rm{o}}^{(z)},{\cal C}_{\rm{o}}^{(z)}} \right).
\end{align} 

\floatname{algorithm}{}
\begin{algorithm} \label{alg2}
\renewcommand{\thealgorithm}{}
\caption{\textbf{Algorithm II}}
\begin{algorithmic}[1]
\State \textbf{Input} $p_i$, ${\bar \gamma _{{{\mathrm{p}}_{_i}}}}$, ${{\bar {\cal I}}_{{{{\rm{p}}}_i}}}$, ${\bar{ \mathcal I}_{{{\mathrm{s}}_i}}}$, ${\bar \upsilon _{{{\mathrm{p}}_i}}}$, ${R_{0,{{{\rm{p}}}_i}}}$, ${{\cal O}_{{p_i}}}$, ${\bar\gamma _{\rm{s}}}$, ${R_{0,{{\rm{s}}}}}$, ${\cal C}_ x^{(z)}$, $p_{\mathrm{s}}^{(0)} = {p_{{\rm{s,max}}}}$, $f\left( {{p_{\rm{s}}},{{\cal C}_x}} \right)={p_{{\rm{out}},{\rm{s}}}}\left( {{p_{\rm{s}}},{{\cal C}_x}} \right)$
\If{${{\cal I}_{\max ,{{{\rm{p}}}_i}}}>{p_j}{{\bar \upsilon }_{{{{\rm{p}}}_j}}}$}
\For {$z=1:k+1$}
\State \textbf{Construct} interval $ ({\cal C}_ x ^{(z-1)} , {\cal C}_x ^{(z)} )$
\State \textbf{Compute} $m = \mathop {\arg \min }\limits_{l \in \{0,1,2\}}   p_{\mathrm{s}}^{( l  )} \left(  \frac{{\cal C}_ x^{(z-1)} +{\cal C}_ x^{(z)}}{2} \right)$ 
\If{ $m =0$}
\State $p_{\rm{o}}^{(z)}\leftarrow {p_{{\rm{s,max}}}}$,   $ \quad \mathcal{C}_\mathrm{o}^{(z)}\leftarrow {\cal C}_ x ^{(z-1)}$
\ElsIf{ ${R_{0,{\rm{s}}}} < {\log _2}\left( {\frac{{{\mu _m}\sqrt {1 + {\Gamma _{{\rm{p}_m}}}} }}{{{\beta _j}{\Gamma _{{\rm{p}_m}}} - {\mu _m}}}} \right)$ is true}
\State $p_{\rm{o}}^{(z)}\leftarrow p_\mathrm{s}^{\left( m \right)}\left({\cal C}_x ^{(z)}\right)$, $\quad \mathcal{C}_{\rm{o}}^{(z)}\leftarrow {\cal C}_x ^{(z)}$
\Else {}
\State $p_{\rm{o}}^{(z)}\leftarrow p_{\mathrm{s}}^{\left( m \right)}\left({\cal C}_x ^{(z-1)}\right)$, $\quad\mathcal{C}_{\rm{o}}^{(z)}\leftarrow {\cal C}_x ^{(z-1)}$
\EndIf 
\EndFor
\State \textbf{Output} $\left( {p_{\rm{s}}^*,{\cal C}_x^*} \right) = \mathop {\arg \min }\limits_{p_{\rm{o}}^{(z)},\;{\cal C}_o^{(z)}} f\left( {p_{\rm{o}}^{(z)},{\cal C}_{\rm{o}}^{(z)}} \right)$
\Else {}
\State \textbf{Output} $\left( {p_{\rm{s}}^*,{\cal C}_x^*} \right) = \left(0,0 \right)$
\EndIf

  \end{algorithmic}
\end{algorithm}
\section{Performance Analysis with perfect direct link  CSI of the secondary user   } 
In this section, we analyze the performance of the underlay spectrum sharing system using improper Gaussian signaling with  perfect CSI of the  SU direct link. Such an assumption is practical since the underlay CR system operates without coordination between the SU and the PU, thus, the SU can estimate its direct link CSI. First, we present analysis of the the SU outage probability with the knowledge of its direct link CSI. Then, we design the SU improper Gaussian signal based on the CSI knowledge to improve the SU outage probability performance. After that, we investigate the benefits of the instantaneous direct link CSI (IDL-CSI) based design versus the average CSI (A-CSI) based design on  both SU and PU.
 
\subsection{Secondary User Outage Probability with Direct Link CSI}
In this subsection, we derive the SU outage probability with perfect SU direct link CSI. First, We can rewrite \eqref{su_out_ineq} as
\begin{align} \label{SU_outage_def}
P_{\rm{out,s}}^{\mathrm{DL}}\left( {{p_{\rm{s}}},{{\cal C}_x}} \right) = \Pr \Big\{ {\Gamma _{\rm{s}}}{{\left( {\Xi  + 1} \right)}^2} &- 2{p_{\rm{s}}}{\gamma _{\rm{s}}}\left( {\Xi  + 1} \right) -  p_{\rm{s}}^2\gamma _{\rm{s}}^2\left( {1 - {\cal C}_x^2} \right) { \ge 0}\Big\},
\end{align}
where $\Xi  = \sum\nolimits_{i = 1}^2 {{p_i}{{\cal I}_{{{{\rm{p}}}_i}}}}$. By solving the quadratic inequality in \eqref{SU_outage_def} with respect to $\Xi$ and after some manipulations, we obtain
\begin{align}\label{su_out_gammas_probability}
P_{\rm{out,s}}^{\mathrm{DL}}\left( {{p_{\rm{s}}},{{\cal C}_x}} \right)=\Pr \left\{ \Xi  \ge \zeta\left( {{p_{\rm{s}}},{{\cal C}_x}} \right)   \right\} = \int_{\zeta\left( {{p_{\rm{s}}},{{\cal C}_x}} \right)}^\infty  {f_{\Xi}\left(z\right)dz},
\end{align}
where $\zeta\left( {{p_{\rm{s}}},{{\cal C}_x}} \right)$ is defined as
\begin{equation}
\zeta\left( {{p_{\rm{s}}},{{\cal C}_x}} \right)=\frac{{{\gamma _{\rm{s}}}\left( {1 - {\cal C}_x^2} \right)}}{{{\Psi _{\rm{s}}}\left( {{p_{\rm{s}}},{{\cal C}_x}} \right)}}-1.
\end{equation}
and $f_{\Xi}\left(z\right)$ is the probability density function (PDF) of the random variable $\Xi$. Since $p_1{{{\cal I}}_{{{{\rm{p}}}_1}}}$ and $p_2{{{\cal I}}_{{{{\rm{p}}}_2}}}$ are independent exponential random variables with mean $p_1{{\bar {\cal I}}_{{{{\rm{p}}}_1}}}$ and $p_2{{\bar {\cal I}}_{{{{\rm{p}}}_2}}}$, respectively,  then $\Xi$ is a hypoexponential random variable with two rate parameters ${1}/{p_1{{\bar {\cal I}}_{{{{\rm{p}}}_1}}}}$ and ${1}/{p_2{{\bar {\cal I}}_{{{{\rm{p}}}_2}}}}$. Hence, according to \cite{ross2010introduction}, its PDF, assuming that ${{{p_1}{{\bar {\cal I}}_{{{{\rm{p}}}_1}}} \neq {p_2}{{\bar {\cal I}}_{{{{\rm{p}}}_2}}}}}$, is found to be
\begin{align}\label{pdf_Xi}
{f_\Xi }\left( {z\left| {{p_1}{{\bar {\cal I}}_{{{{\rm{p}}}_1}}} \neq {p_2}{{\bar {\cal I}}_{{{{\rm{p}}}_2}}}} \right.} \right) = \sum\limits_{{  i = 1 \atop
j \ne i }}^2 {\frac{{\exp \left( { - \frac{z}{{{p_i}{{\bar {\cal I}}_{{{{\rm{p}}}_i}}}}}} \right)}}{{{p_i}{{\bar {\cal I}}_{{{{\rm{p}}}_i}}} - {p_j}{{\bar {\cal I}}_{{{{\rm{p}}}_j}}}}}} \mathds{1}_{\left[ 0, \infty \right)}\left( z \right).
\end{align} 
For ${{{p_1}{{\bar {\cal I}}_{{{{\rm{p}}}_1}}} = {p_2}{{\bar {\cal I}}_{{{{\rm{p}}}_2}}}}}$, the PDF of $\Xi$ reduces to
\begin{align}
{f_\Xi }\left( {z\left| {{p_1}{{\bar {\cal I}}_{{{{\rm{p}}}_1}}} = {p_2}{{\bar {\cal I}}_{{{{\rm{p}}}_2}}}} \right.} \right)=\frac{z\exp \left( { - \frac{z}{{{p_1}{{\bar {\cal I}}_{{{{\rm{p}}}_1}}}}}} \right)}{{{{\left( {{p_1}{{\bar {\cal I}}_{{{{\rm{p}}}_1}}}} \right)}^2}}}\mathds{1}_{\left[ 0, \infty \right)}\left( z \right),
\end{align}
which represents the Erlang distribution with parameters, shape $k=2$ and rate $\lambda=1/{{{{ {{p_1}{{\bar {\cal I}}_{{{{\rm{p}}}_1}}}} }}}}$~\cite{forbes2011statistical}.

From \eqref{su_out_gammas_probability}, we note that $P_{\rm{out,s}}^{\mathrm{DL}}\left( {{p_{\rm{s}}},{{\cal C}_x}} \right)$ attains its maximum value of unity if $\zeta\left( {{p_{\rm{s}}},{{\cal C}_x}} \right) \leq 0$. Thus, the SU outage probability in this scenario can be rewritten~as
\begin{align} \label{SU_out_DL}
P_{{\rm{out,s}}}^{{\rm{DL}}}\left( {{p_{\rm{s}}},{{\cal C}_x}} \right) = \left\{ {\begin{array}{*{20}{c}}
1,& \;{{\gamma _{\rm{s}}} \le \frac{{{\Psi _{\rm{s}}}\left( {{p_{\rm{s}}},{{\cal C}_x}} \right)}}{{1 - {\cal C}_x^2}}}\\  \\
{P_{{\rm{out,s}}}^{{\rm{DL - T}}}\left( {{p_{\rm{s}}},{{\cal C}_x}} \right)},& \;\;{{\gamma _{\rm{s}}} > \frac{{{\Psi _{\rm{s}}}\left( {{p_{\rm{s}}},{{\cal C}_x}} \right)}}{{1 - {\cal C}_x^2}}}
\end{array}} \right.,
\end{align}
where we obtain 100$\%$ outage of the SU transmission for highly faded SU direct link channel regardless of the PU interference level on the SU, thus no-transmission should be adopted at the SU side. On the other hand, when the SU has a good direct link channel, it transmits. The transmission condition is defined according to \eqref{SU_out_DL} 
\begin{align} \label{trans_cond}
{{\gamma _{\rm{s}}} > \frac{{{\Psi _{\rm{s}}}\left( {{p_{\rm{s}}},{{\cal C}_x}} \right)}}{{1 - {\cal C}_x^2}}}.
\end{align}
Then, we obtain the corresponding outage probability under perfect knowledge of $\gamma_{\rm{s}}$ as
\footnotesize
\begin{align} \label{SU_out_trans}
P_{\rm{out,s}}^{\mathrm{DL-T}}\left( {{p_{\rm{s}}},{{\cal C}_x}} \right)= \left\{ {\begin{array}{*{20}{c}}
\sum\limits_{i = 1\atop
j \ne i}^2 {\frac{{{p_i}{{\bar {\cal I}}_{{{{\rm{p}}}_i}}}\exp \left( { - \frac{\zeta\left( {{p_{\rm{s}}},{{\cal C}_x}} \right)}{{{p_i}{{\bar {\cal I}}_{{{{\rm{p}}}_i}}}}} } \right)}}{{{p_i}{{\bar {\cal I}}_{{{{\rm{p}}}_i}}} - {p_j}{{\bar {\cal I}}_{{{{\rm{p}}}_j}}}}}},&{{{p_1}{{\bar {\cal I}}_{{{{\rm{p}}}_1}}} \neq {p_2}{{\bar {\cal I}}_{{{{\rm{p}}}_2}}}}} \\  \\
\left( {1 + \frac{{\zeta \left( {{p_{\rm{s}}},{{\cal C}_x}} \right)}}{{{p_1}{{\bar {\cal I}}_{{{\rm{p}}_1}}}}}} \right)\exp \left( { - \frac{{\zeta \left( {{p_{\rm{s}}},{{\cal C}_x}} \right)}}{{{p_1}{{\bar {\cal I}}_{{{\rm{p}}_1}}}}}} \right),& {{{p_1}{{\bar {\cal I}}_{{{{\rm{p}}}_1}}} = {p_2}{{\bar {\cal I}}_{{{{\rm{p}}}_2}}}}}  
\end{array}} \right..
\end{align}
\normalsize
According to the availability of SU CSI, the SU will not be allowed to access the spectrum if \eqref{trans_cond} is not valid. On the other hand, if the transmission condition is valid, i.e., \eqref{trans_cond}, then SU outage is governed not only by $\gamma_{\rm{s}}$, but also by the PU interference links on the SU. 
%
%

For the proper case, i.e., $\mathcal{C}_x = 0$, the SU outage probability, when transmitting, reduces to
\footnotesize
\begin{align}
P_{{\rm{out,s}}}^{{\rm{DL-T}}}\left( {{p_{\rm{s}}},0} \right)= \left\{ {\begin{array}{*{20}{c}}
\sum\limits_{i = 1\atop
j \ne i}^2 {\frac{{{p_i}{{\bar {\cal I}}_{{{\rm{p}}_i}}}\exp \left( { - \frac{{\frac{{{p_{\rm{s}}}{\gamma _{\rm{s}}}}}{2^{R_{0,\rm{s}}}-1} - 1}}{{{p_i}{{\bar {\cal I}}_{{{\rm{p}}_i}}}}}} \right)}}{{{p_i}{{\bar {\cal I}}_{{{\rm{p}}_i}}} - {p_j}{{\bar {\cal I}}_{{{\rm{p}}_j}}}}}},&{{{p_1}{{\bar {\cal I}}_{{{{\rm{p}}}_1}}} \neq {p_2}{{\bar {\cal I}}_{{{{\rm{p}}}_2}}}}}\\  \\
\left( {1 + \frac{{\frac{{{p_{\rm{s}}}{\gamma _{\rm{s}}}}}{2^{R_{0,\rm{s}}}-1} - 1}}{{{p_1}{{\bar {\cal I}}_{{{\rm{p}}_1}}}}}} \right)\exp \left( { - \frac{{\frac{{{p_{\rm{s}}}{\gamma _{\rm{s}}}}}{2^{R_{0,\rm{s}}}-1} - 1}}{{{p_1}{{\bar {\cal I}}_{{{\rm{p}}_1}}}}}} \right),& {{p_1}{{\bar {\cal I}}_{{{{\rm{p}}}_1}}} = {p_2}{{\bar {\cal I}}_{{{{\rm{p}}}_2}}}}
\end{array}} \right..
\end{align} 
\normalsize
while for maximally improper case, i.e., $\mathcal{C}_x = 1$, it yields
\footnotesize
\begin{align}
\mathop {\lim }\limits_{{{\cal C}_x} \to 1} P_{{\rm{out,s}}}^{{\rm{DL-T}}}\left( {{p_{\rm{s}}},{{\cal C}_x}} \right)= \left\{ {\begin{array}{*{20}{c}}
\sum\limits_{i = 1\atop j \ne i}^2 {\frac{{{p_i}{{\bar {\cal I}}_{{{\rm{p}}_i}}}\exp \left( { - \frac{{\frac{{2{p_{\rm{s}}}{\gamma _{\rm{s}}}}}{{{\Gamma _{\rm{s}}}}} - 1}}{{{p_i}{{\bar {\cal I}}_{{{\rm{p}}_i}}}}}} \right)}}{{{p_i}{{\bar {\cal I}}_{{{\rm{p}}_i}}} - {p_j}{{\bar {\cal I}}_{{{\rm{p}}_j}}}}}},& {{{p_1}{{\bar {\cal I}}_{{{{\rm{p}}}_1}}} \neq {p_2}{{\bar {\cal I}}_{{{{\rm{p}}}_2}}}}}\\  \\
\left( {1 + \frac{{\frac{{2{p_{\rm{s}}}{\gamma _{\rm{s}}}}}{{{\Gamma _{\rm{s}}}}} - 1}}{{{p_1}{{\bar {\cal I}}_{{{\rm{p}}_1}}}}}} \right)\exp \left( { - \frac{{\frac{{2{p_{\rm{s}}}{\gamma _{\rm{s}}}}}{{{\Gamma _{\rm{s}}}}} - 1}}{{{p_1}{{\bar {\cal I}}_{{{\rm{p}}_1}}}}}} \right),&{{p_1}{{\bar {\cal I}}_{{{{\rm{p}}}_1}}} = {p_2}{{\bar {\cal I}}_{{{{\rm{p}}}_2}}}}
\end{array}} \right..
\end{align}
\normalsize
\subsection{Design of Secondary User Signal Parameters Based on Perfect Direct Link CSI }
In this subsection, we aim to design the SU signal parameters, i.e.,  $p_{\mathrm{s}}$ and $\mathcal{C}_x$, in the improper Gaussian signaling case, and merely  $p_{\mathrm{s}}$, in the proper case, in order to satisfy the PU QoS and the boundary values for these parameters. Therefore, we solve the optimization problem for the proper proper case
\begin{align} \label{opt_prob_proper_inst}
 & \mathop {\min }\limits_{{p_{\rm{s}}}} \quad P_{\rm{out,s}}^{\mathrm{DL}}\left( {{p_{\rm{s}}},0} \right) \hspace{0.25cm} \nonumber \\
& \hspace{0.1cm}\mathrm{s. \; t.} \quad  {P_{\mathrm{out},{{\mathrm{p}}_i}}}\left( {{p_{\mathrm{s}}},0} \right) \le {{\cal O}_{{p_i}}}, \nonumber \\
&\quad \quad \quad 0 < {p_{\mathrm{s}}} \le {p_{{\mathrm{s,max}}}},
\end{align}
and the following optimization problem for the improper case,
\begin{align} \label{opt_prob_improper_inst}
&\mathop {\min }\limits_{{p_{\rm{s}}},{{\cal C}_x}} \quad P_{\rm{out,s}}^{\mathrm{DL}}\left( {{p_{\rm{s}}},{{\cal C}_x}} \right) \hspace{0.25cm} \nonumber \\
&\hspace{0.1cm}\mathrm{s. \; t. \;} \quad {P_{\mathrm{out},{{{\rm{p}}}_i}}^{\rm{UB}}}\left( {{p_{\rm{s}}},\;{\cal C}_x} \right) \le {{\cal O}_{{p_i}}}, \nonumber \\
& \quad \quad \quad \;0 \le {p_{\rm{s}}} \le {p_{{\rm{s,max}}}}, \nonumber \\
& \quad \quad \quad \;0 \leq {{\cal C}_x} \le 1.
\end{align}

 First, for the proper Gaussian signaling case, one can show that $P_{\rm{out,s}}^{\mathrm{DL}}\left( {{p_{\rm{s}}},0} \right)$ is monotonically decreasing in $p_\mathrm{s}$, thus the optimal minimum SU outage probability is achieved by allocating the SU power to the minimum of the three functions in \eqref{ps_proper_avg_case} as long as the transmission condition is valid, i.e.,  \eqref{trans_cond}, otherwise the SU should stay  silent because a unity SU outage probability is expected to be achieved. Therefore, as a modification of the previous section, the SU needs to check the validity of transmission condition in \eqref{trans_cond} at $\mathcal{C}_x=0$, to stay silent or to transmit with a power that is computed from Algorithm I.

For the improper case, if $p_{\rm{s}}= p_{\rm{s}}^{\left( m \right)}, m = \{1,2\}$ in \eqref{ps_min}, then the SU outage probability in the transmission scenario, i.e., \eqref{trans_cond} is valid, can be written in terms of  $ \mathcal{C}_x$ as
\footnotesize
\begin{align}\label{p_out_s_DL_Cx}
P_{{\rm{out,s}}}^{\mathrm{DL-T}}\left( \mathcal{C}_x  \right)= \left\{ {\begin{array}{*{20}{c}}
\sum\limits_{i = 1\atop j \ne i}^2 {{\cal K}{_{i,j}}\exp \left( { - \frac{{{\cal Y}{{\cal G}_m}\left( {{\gamma _{\rm{s}}}} \right) - 1}}{{{p_i}{{\bar {\cal I}}_{{{\rm{p}}_i}}}}}} \right)}, &{{{p_1}{{\bar {\cal I}}_{{{{\rm{p}}}_1}}} \neq {p_2}{{\bar {\cal I}}_{{{{\rm{p}}}_2}}}}}  \\  \\
\left( {1 + \frac{{{\cal Y}{{\cal G}_m}\left( {{\gamma _{\rm{s}}}} \right) - 1}}{{{p_1}{{\bar {\cal I}}_{{{\rm{p}}_1}}}}}} \right)\exp \left( { - \frac{{{\cal Y}{{\cal G}_m}\left( {{\gamma _{\rm{s}}}} \right) - 1}}{{{p_1}{{\bar {\cal I}}_{{{\rm{p}}_1}}}}}} \right),& {{{p_1}{{\bar {\cal I}}_{{{{\rm{p}}}_1}}} = {p_2}{{\bar {\cal I}}_{{{{\rm{p}}}_2}}}}} 
\end{array}} \right.,
\end{align} \normalsize
where ${\cal K}{_{i,j}} = {p_i}{{\bar {\cal I}}_{{{\rm{p}}_i}}}/\left( {{p_i}{{\bar {\cal I}}_{{{\rm{p}}_i}}} - {p_j}{{\bar {\cal I}}_{{{\rm{p}}_j}}}} \right)$. Thus, ${P_\mathrm{out,s}}\left( {{{\cal C}_x}}\right)$ is monotonically decreasing in $\mathcal{C}_x$, when the condition in \eqref{improper_condition} is true\footnote{See Appendix C for the proof.}. Moreover, if $p_{\rm{s}}={p_{{\rm{s,max}}}}$ in \eqref{ps_min}, ${P_\mathrm{out,s}}\left( {{{\cal C}_x}}\right)$ is monotonically increasing in $\mathcal{C}_x$ which can be proven similar to the proof in Appendix C. Thus, same steps are applied to obtain the optimal $( {p_{\rm{o}}^{(z)},{\cal C}_{\rm{o}}^{(z)}} )$ pairs in each of the intervals defined by the boundary points ${\cal C}_ x^{(z)}$, i.e., steps 6 to 12 in Algorithm II. At the end, we calculate the global optimal pair $\left( {p_{\rm{s}}^*,{\cal C}_x^*} \right)$ from
\begin{align}\label{opt_prob_global_inst}
\left( {p_{\rm{s}}^*,{\cal C}_x^*} \right) = \mathop {\arg \min } {p_{{\rm{out}},{\rm{s}}}^{\rm{DL-T}}}\left( {p_{\rm{o}}^{(z)},{\cal C}_{\rm{o}}^{(z)}} \right), 
\end{align}  
which can be further simplified and formulated as\footnote{See Appendix D for the proof.}
\begin{align}\label{opt_prob_equivalent}
\left( {p_{\rm{s}}^*,{\cal C}_x^*} \right) = \mathop {\arg \min } \frac{{  {1 - {{\left( {{\cal C}_o^{(z)}} \right)}^2}}  }}{{{\Psi _{\rm{s}}}\left( {p_o^{(z)},{\cal C}_o^{(z)}} \right)}}.
\end{align}   
It is clear from \eqref{opt_prob_equivalent} that the optimal parameters are independent of ${{\gamma _{\rm{s}}}}$. Thus, the availability of the direct link CSI information determines only whether the SU can transmit or not as can be seen from \eqref{SU_out_DL} and \eqref{trans_cond}. 
  As a result, by knowing the direct link CSI, the SU is able to save some of its transmit power and hence boost its average EE as will be discussed in the following~subsections.
\subsection{Benefits of SU Direct Link CSI}
In this subsection, we investigate the benefits obtained by the SU through perfect CSI knowledge of the SU direct link. 
\subsubsection{SU power saving}
In the improper design of the spectrum sharing system based on the average statistics of the channel coefficients in Section IV, it is clear that the SU always transmits if the maximum INR for user $i$ exceeds ${p_j}{{\bar \upsilon }_{{{{\rm{p}}}_j}}}$. On the contrary, based on perfect CSI of the SU direct link, the SU stays silent if the transmission condition  \eqref{trans_cond} is not satisfied because it yields a unity SU outage probability. Thus, the probability of the power saving event is expressed~as
\begin{align}
{P}_{\rm{saving}}&=\Pr \left\{ {{\gamma _{\rm{s}}} \le \frac{{{\Psi _{\rm{s}}}\left( {{p_{\rm{s}}},{{\cal C}_x}} \right)}}{{\left( {1 - {\cal C}_x^2} \right)}}} \right\} =1 - \exp \left( { - \frac{{{\Psi _{\rm{s}}}\left( {{p_{\rm{s}}},{{\cal C}_x}} \right)}}{{{{\bar \gamma }_{\rm{s}}}\left( {1 - {\cal C}_x^2} \right)}}} \right),
\end{align} 
which represents also the portion of time that the SU does not consume energy and and stays~idle. 
\subsubsection{Improving the SU average energy efficiency}
The improper Gaussian signaling design tends to use more  power to improve the SU design and relieve its impact on the PU by increasing the circularity coefficient, as can be concluded from the monotonically decreasing characteristics of the $P_{\mathrm{out,s}}$ in $\mathcal{C}_x$ discussed in Sections (IV-B and V-B). Since the SU CSI offers some power saving advantages, then the average EE of the SU system is expected to improve in this scenario. In the following, we analyze the average EE of the SU system when using the SU direct link CSI comparing to the design case based on the average statistics of CSI.


The EE of the SU system is defined as the ratio between the data that is successfully delivered to the receiver and the corresponding total energy consumption \cite{amin2014novel}. Firstly, for the average statistics based design, the SU always transmits and the average EE of the SU is expressed as
\begin{align} \label{EE_avg_design}
\bar \eta _{{\rm{EE}}}^{{\rm{Avg}}} =\frac{{{R_{0,{\rm{s}}}}\left( {1 - {p_{{\rm{out,s}}}}\left( {{p_{\rm{s}}},{{\cal C}_x}} \right)} \right)}}{{\kappa {p_{\rm{s}}} + {p_c}}},
\end{align} 
where $\kappa$ is the reciprocal power amplifier efficiency and $p_\mathrm{c}$ is the circuits power consumption.

Secondly, for the case of known direct link CSI, the average EE is defined when the SU is allowed to transmit, i.e., ${\gamma _{\rm{s}}} \geq {{{\Psi _{\rm{s}}}\left( {{p_{\rm{s}}},{{\cal C}_x}} \right)}}{/{\left( {1 - {\cal C}_x^2} \right)}}$, as
\begin{align}\label{EE_avg_SU}
\bar \eta _{{\rm{EE}}}^{{\rm{DL}}} = {\mathbb{E}_{{\gamma _{\rm{s}}}}}\left\{ {{\eta _{{\rm{EE}}}}\left( {{\gamma _{\rm{s}}}\left| {{\gamma _{\rm{s}}}} \right. > \frac{{{\Psi _{\rm{s}}}\left( {{p_{\rm{s}}},{{\cal C}_x}} \right)}}{{1 - {\cal C}_x^2}}} \right)} \right\},
\end{align} 
where ${\eta _{{\rm{EE}}}}$ represents the EE of the SU at a specific direct link CSI, which is defined as
\begin{align}
{\eta _{{\rm{EE}}}}\left( {{\gamma _{\rm{s}}}} \right) = \frac{{{R_{0,{\rm{s}}}}\left( {1 - p_{{\rm{out,s}}}^{{\rm{DL}}}\left( {{p_{\rm{s}}},{{\cal C}_x}} \right)} \right)}}{{{P_{\rm{T}}}\left( {{\gamma _{\rm{s}}}} \right)}},
\end{align} 
where ${P_{\rm{T}}}\left( {{\gamma _{\rm{s}}}} \right)$ is the total power used to deliver the data and defined as 
\begin{align}
{P_{\rm{T}}}\left( {{\gamma _{\rm{s}}}} \right) = \left( {\kappa {p_{\rm{s}}} + {p_{\rm{c}}}} \right)\mathds{1}_{\left[ \frac{{{\Psi _{\rm{s}}}\left( {{p_{\rm{s}}},{{\cal C}_x}} \right)}}{{\left( {1 - {\cal C}_x^2} \right)}},\infty\right)}\left(\gamma_{\rm{s}}\right).
\end{align} 
Here, we have assumed that the SU transmitter consumes negligible power in the idle/sleep mode. From \eqref{EE_avg_SU}, the SU average EE can be written as
\begin{align}\label{EE_avg_SU_integral}
\bar \eta _{{\rm{EE}}}^{{\rm{DL}}} = \frac{{\int\limits_{\frac{{{\Psi _{\rm{s}}}\left( {{p_{\rm{s}}},{{\cal C}_x}} \right)}}{{1 - {\cal C}_x^2}}}^\infty  {{\eta _{{\rm{EE}}}}\left( {{\gamma _{\rm{s}}}} \right)\exp \left( { - \frac{x}{{{{\bar \gamma }_{\rm{s}}}}}} \right)} dx}}{{{{\bar \gamma }_{\rm{s}}}\left( {1 - {P_{{\rm{saving}}}}} \right)}}.
\end{align} 
After evaluating the integral in \eqref{EE_avg_SU_integral}, we obtain 
\begin{align}\label{EE_avg_SU_expression}
\bar \eta _{{\rm{EE}}}^{{\rm{DL}}} =\frac{{{R_{0,{\rm{s}}}}}}{{\left( {\kappa {p_{\rm{s}}} + {p_c}} \right)\prod\limits_{j = 1}^2 {\left( {{p_j}{{\overline {\cal I} }_{{{\rm{p}}_j}}}\frac{{{\Psi _{\rm{s}}}\left( {{p_{\rm{s}}},{{\cal C}_x}} \right)}}{{{{\bar \gamma }_{\rm{s}}}\left( {1 - {\cal C}_x^2} \right)}} + 1} \right)} }}.
\end{align}
From  \eqref{EE_avg_design} and \eqref{EE_avg_SU_expression}, we compute the improvement in the average EE, $\mathpzc{E}$, after simplification as
\begin{align}
\mathpzc{E}= \frac{ \bar \eta _{{\rm{EE}}}^{{\rm{DL}}}}{\bar \eta _{{\rm{EE}}}^{{\rm{Avg}}}} = \frac{1}{{1 - {P_{{\rm{saving}}}}} }.
\end{align}

\subsubsection{PU Outage Probability Enhancement}
As discussed in the previous 2 points, the SU can make use of the CSI of  direct link in order to save its transmit power and enhance the average EE performance. Moreover, when the SU detects such outage events and decides to abandon its transmission, the interference impact on the PU will be significantly alleviated and hence, the outage performance of the PU will be improved as follows
\begin{align}
&P_{{\rm{out,}}{{\rm{p}}_i}}^{\mathrm{DL}}\left( {{p_{\rm{s}}},{{\cal C}_x}} \right) = {P_{{\rm{saving}}}}P_{{\rm{out,}}{{\rm{p}}_i}} \left( {0,0} \right) + \left( {1 - {P_{{\rm{saving}}}}} \right)P_{{\rm{out,}}{{\rm{p}}_i}} \left( {{p_{\rm{s}}},{{\cal C}_x}} \right)\nonumber \\
& =P_{{\rm{out,}}{{\rm{p}}_i}}\left( {{p_{\rm{s}}},{{\cal C}_x}} \right) - {P_{{\rm{saving}}}}\underbrace {\left( {P_{{\rm{out,}}{{\rm{p}}_i}}\left( {{p_{\rm{s}}},{{\cal C}_x}} \right) - P_{{\rm{out,}}{{\rm{p}}_i}}\left( {0,0} \right)} \right)}_{ \mathop  \geq \limits^{(a)} 0}  \le P_{{\rm{out,}}{{\rm{p}}_i}}\left( {{p_{\rm{s}}},{{\cal C}_x}} \right),
\end{align}   
where $(a)$ follows from the fact that interference free system performance is better than systems subjected to any type of interference. Thus, SU direct link channel knowledge  provides a protection for the PU performance that may be used by other cognitive users in the same system to access the spectrum without deteriorating the PU QoS.
\section{Numerical Results}
In this section, we present some numerical examples and simulations that validate the introduced analysis and investigate the benefits of employing improper Gaussian signaling in spectrum sharing with FD PU. First, we compare the proposed PU outage probability upper bound  with the exact expression that is computed numerically. Thereafter, we exploit these bounds to design the SU signal parameters in order to minimize the SU outage probability while maintaining certain QoS requirements for the PU based on the average CSI and also examine the effect of RSI channel of the employment of improper Gaussian signaling. Moreover, we investigate the proposed system design in case of the perfect knowledge of the direct link CSI and verify the benefits that the SU can attain in terms of power saving and hence average EE improvement. Furthermore, we provide numerical simulations of  the benefit that is achieved at the PU side. 

Throughout this Section, we use the following general system parameters for all examples, unless otherwise specified. For the PU nodes, we assume ${R_{0,{{{\rm{p}}}_i}}}=0.5$ b/s/Hz with a maximum power budget $p_i=1 \; \rm{W}$. The communications channels are characterized as, ${\bar \gamma _{{{\mathrm{p}}_{_i}}}}=25$ dB, ${{\bar {\cal I}}_{{{{\rm{p}}}_i}}}=3$ dB, ${\bar \upsilon _{{{\mathrm{p}}_i}}}=5$ dB. We assume that the required PU outage probability threshold ${{\cal O}_{{p_i}}}=0.01$. The SU is assumed to target ${R_{0,{{\rm{s}}}}}=0.5$ b/s/Hz using $p_{\rm{s,max}}=1 \; \rm{W}$. The SU channels' parameters are assumed to be ${{\bar {\cal I}}_{{{\rm{s}}_i}}}=13$ dB and ${\bar \gamma _{{{\mathrm{s}}}}}=20$ dB. For the IDL-CSI based design, we used $10^6$ independent channel realizations for the direct link of the SU in the simulation.

\begin{figure}[!t]
\centering
\includegraphics[width=9cm]{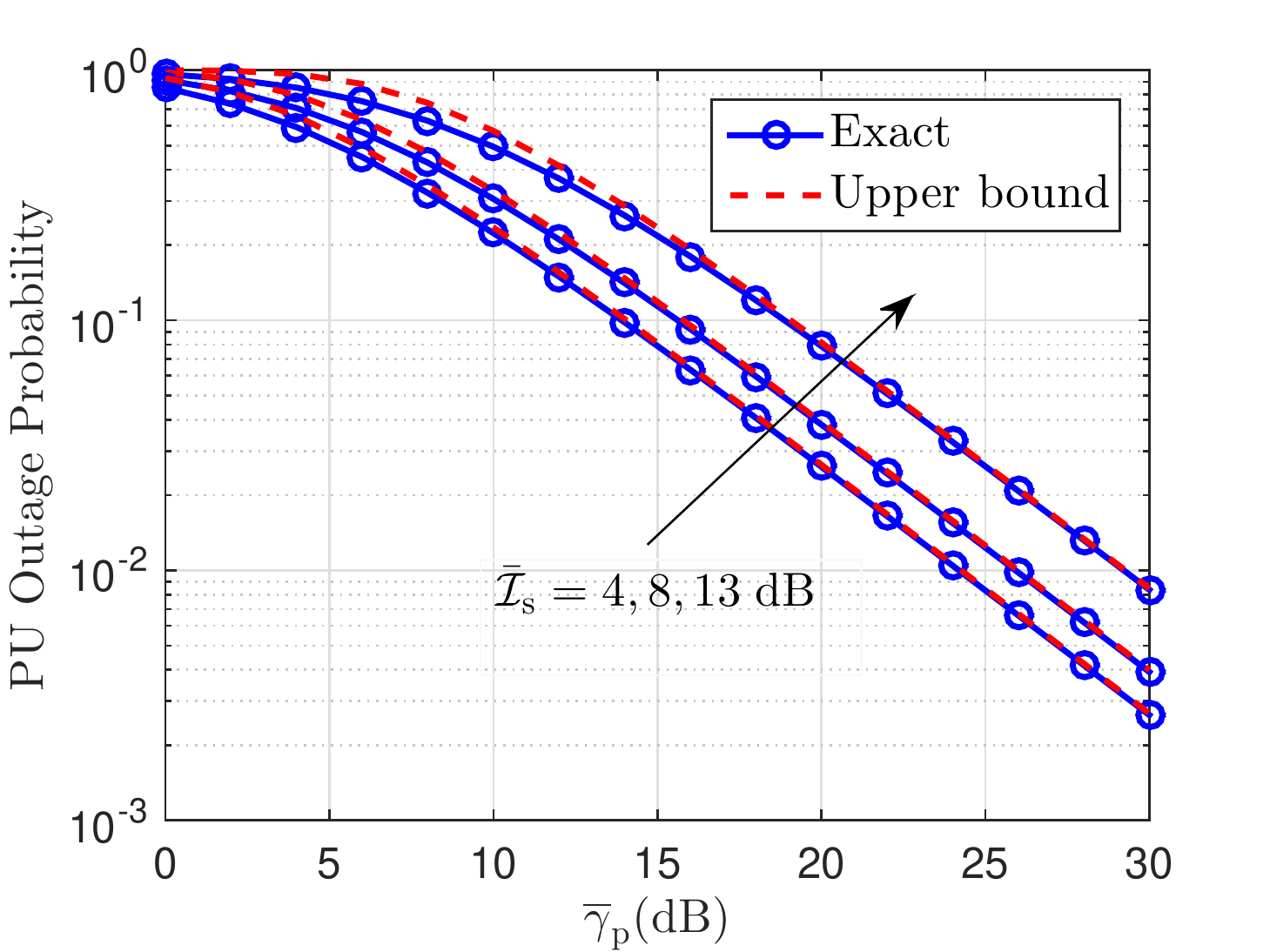}
\caption{A comparison between the exact PU outage probability and the upper bound versus ${\bar \gamma _{{{\mathrm{p}}}}}$ for ${{\bar {\cal I}}_{{{\rm{s}}}}}=4, 8, 13 \; \mathrm{dB}$.}
\label{SimEx1}
\end{figure}

\textit{Example 1:} This example compares the upper bound computed from \eqref{pu_outage_ub} with the exact value computed by evaluating the expectations in \eqref{pu_integral} numerically. We assume ${\bar \gamma _{{{\mathrm{p}}_{_i}}}}={\bar \gamma _{{{\mathrm{p}}}}}$, $\mathcal{C}_x=0.5$ and ${{\bar {\cal I}}_{{{\rm{s}}_i}}}={{\bar {\cal I}}_{{{\rm{s}}}}}=4, 8, 13$ dB. As shown in Fig. \ref{SimEx1}, the upper bound is tight to the exact outage probability for different ${{\bar {\cal I}}_{{{\rm{s}}}}}$. Similar results are observed for different ${R_{0,{{{\rm{p}}}_i}}}$.

\begin{figure}[!t]
\centering
\includegraphics[width=9cm]{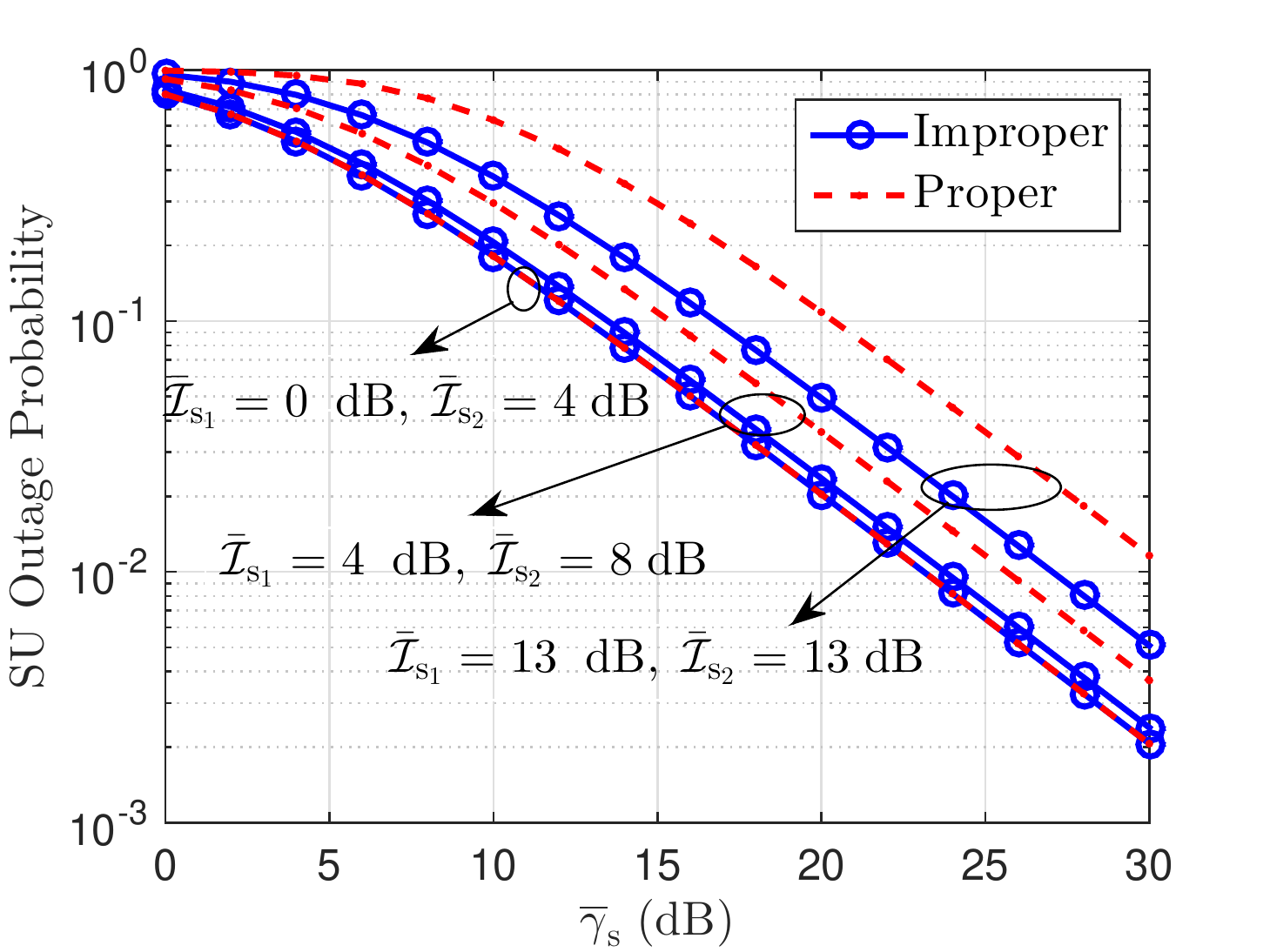}
\caption{SU outage probability for proper and improper Gaussian signaling versus ${\bar \gamma _{{{\mathrm{s}}}}}$ for $\left(\bar {{\cal I}}_{{{\rm{s}}_1}},\bar {{\cal I}}_{{{\rm{s}}_2}}\right)=\left(0,4\right),\left(4,8\right),\left(13,13\right)\;\rm{dB}$ pairs.}
\label{SimEx2}
\end{figure}

\textit{Example 2:} In this example, we inspect the benefits of designing the improper Gaussian signaling for SU over the conventional proper Gaussian signaling design. We assume the pair $\left(\bar {{\cal I}}_{{{\rm{s}}_1}},\bar {{\cal I}}_{{{\rm{s}}_2}}\right)$ has has the following values, $\left(0,4\right)$ dB, $\left(4,8\right)$ dB and $\left(13,13\right)$ dB. The proper design is based on Algorithm I. For the improper design, we first obtain the distinct intersection points, if exist, and sort them in $\mathcal{C}_x^{z}$, then we   obtain the optimal pair $\left(p_{{\rm{s}}}^*,\mathcal{C}_x^*\right)$ by applying Algorithm II. Fig. \ref{SimEx2} shows the SU outage probability versus ${\bar \gamma _{{{\mathrm{s}}}}}$ for different pairs of ${{\bar {\cal I}}_{{{\rm{s}_i}}}}$. For $\left(\bar {{\cal I}}_{{{\rm{s}}_1}},\bar {{\cal I}}_{{{\rm{s}}_2}}\right)=\left(0,4\right)$ dB, there is no gain from using improper signaling. In this case, the interference channel is week, which allow the SU with proper signaling to improve its performance (minimize its outage probability) by increasing the transmitted power and employing the maximum budget. As we observed from the improper design, $p_\mathrm{s}$ tends to increase with $\mathcal{C}_x$  as can be seen in \eqref{ps_improper_max}, but since $p_\mathrm{s}(0) \simeq p_{\mathrm{s,max}}$, then the improper solution reduces approximately to the proper design. As the SU interference channels $\bar {{\cal I}}_{{{\rm{s}}_i}}$ become stronger, proper signaling system uses less power to meet PU QoS requirement while improper signaling can use more power to improve its outage probability performance while compensating for its interference impact on the PU by increasing the circularity coefficient. Fig. \ref{SimEx2} shows a $1.5-3.5$ dB improvement resulting from adopting improper Gaussian signaling.

\begin{figure}[!t]
\centering
\includegraphics[width=9cm]{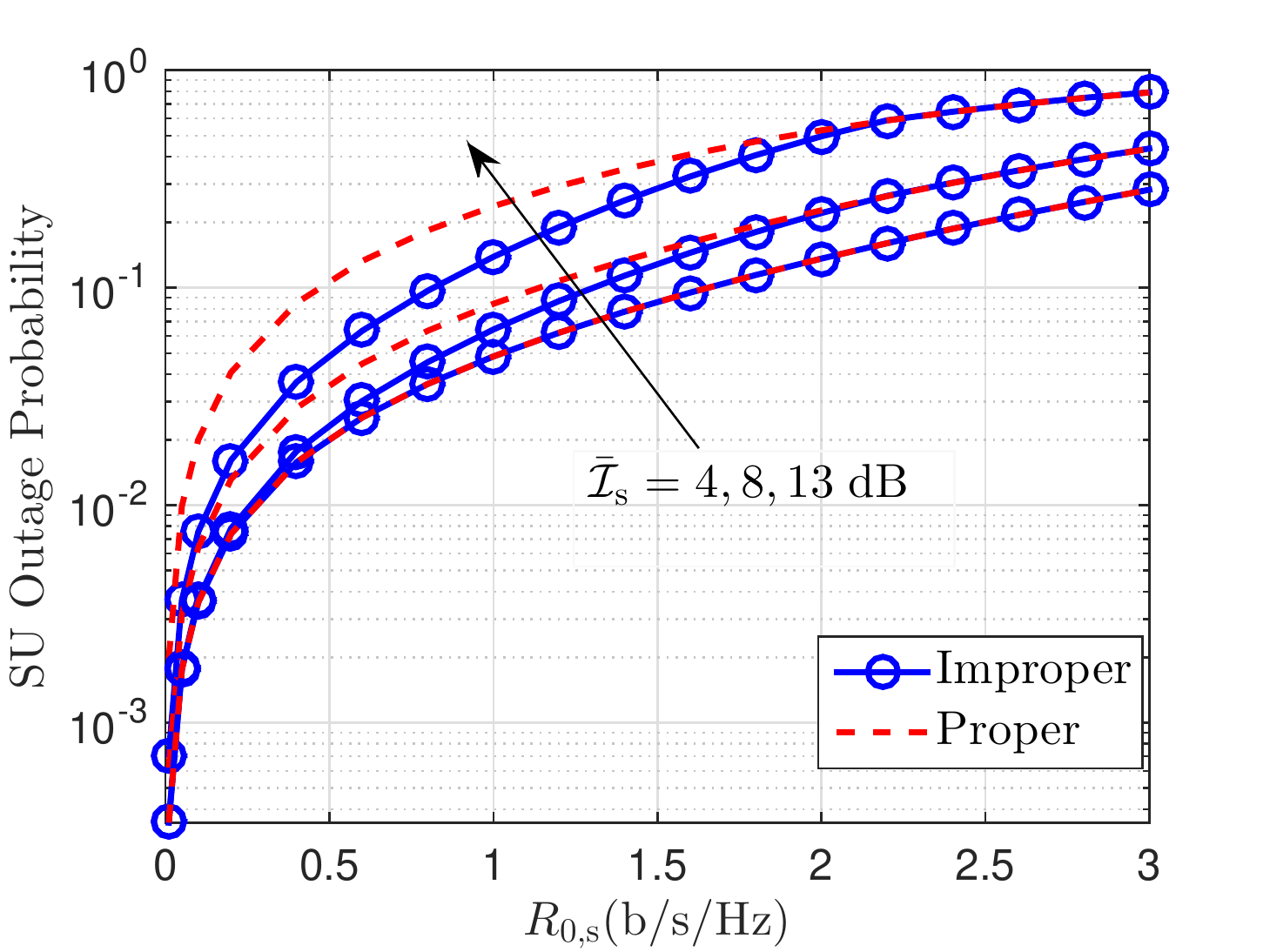}
\caption{SU outage probability for proper and improper Gaussian signaling versus ${R_{0,{{\rm{s}}}}}$ for ${{\bar {\cal I}}_{{{\rm{s}}}}}=4, 8, 13 \; \mathrm{dB}$.}
\label{SimEx3}
\end{figure}

\textit{Example 3:} Fig. \ref{SimEx3} plots the SU outage probability versus different SU target rates ${R_{0,{{\rm{s}}}}}$. We assume that $\bar {{\cal I}}_{{{\rm{s}}_i}}=\bar {{\cal I}}_{{{\rm{s}}}}=4, 8, 13$ dB. Similar to the previous example, it is clear that improper Gaussian signaling system achieves superior performance that the proper one when the SU interference channel to the PU is strong. However, at high SU target rates, there is no gain from employing improper signaling as can be deduced from the condition in \eqref{improper_condition}.   

\begin{figure}[!t]
\centering
\includegraphics[width=9cm]{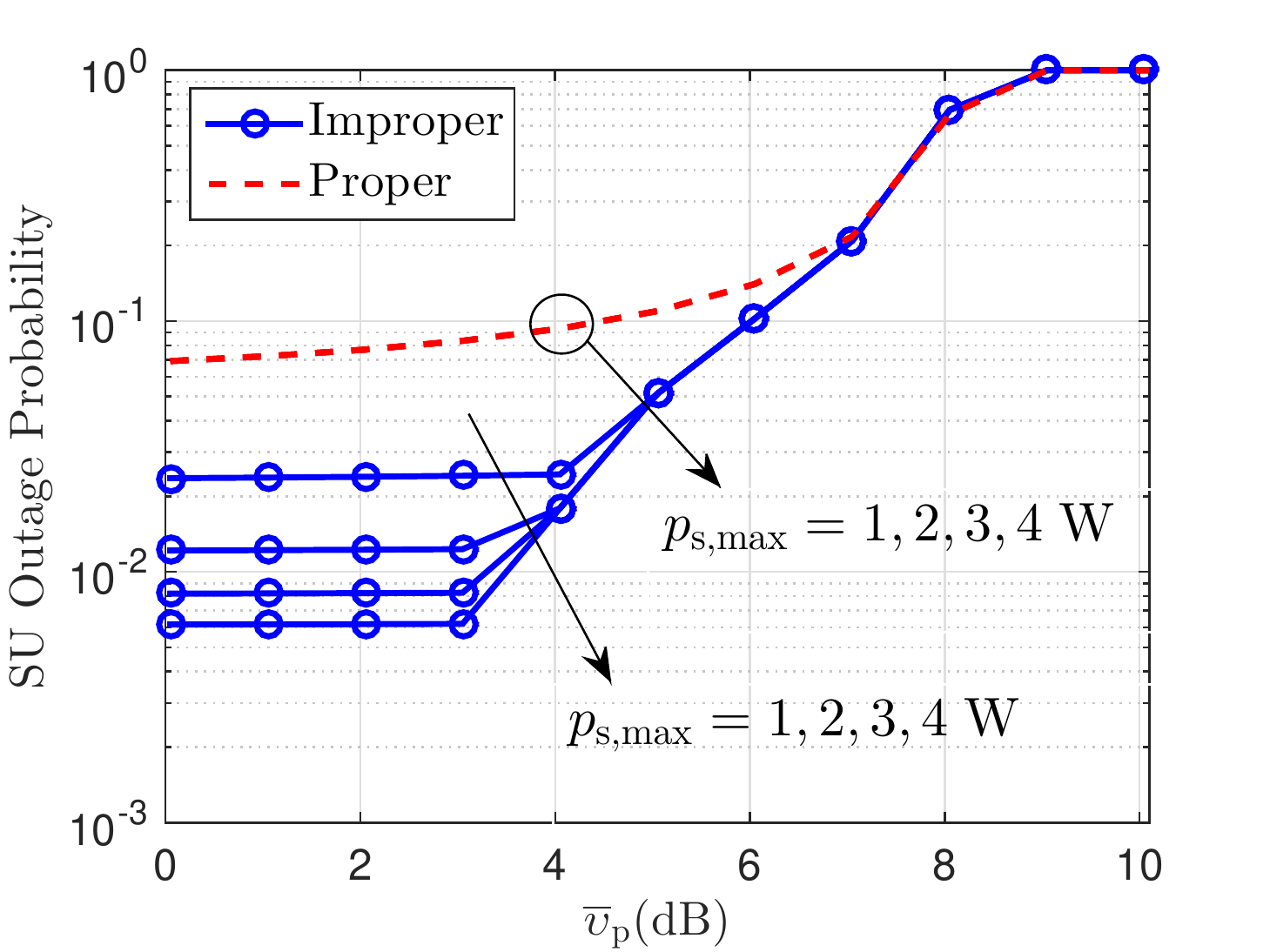}
\caption{SU outage probability for proper and improper Gaussian signaling versus ${\bar \upsilon _{{{\mathrm{p}}}}}$ for $p_{\rm{s,max}}=1,2,3,4\;\rm{W}$.}
\label{SimEx4}
\end{figure}

\textit{Example 4:} This example investigates the impact of RSI-CNR in limiting the CR operation and compares between its effect on both proper and improper Gaussian signaling based systems. We assume ${\bar \upsilon _{{{\mathrm{p}_i}}}}={\bar \upsilon _{{{\mathrm{p}}}}}$. For this purpose, we plot the SU outage probability versus ${\bar \upsilon _{{{\mathrm{p}}}}}$ for different values of $p_{\rm{s,max}}$ in Fig. \ref{SimEx4}. We observe that improper Gaussian signaling achieves better performance than the proper Gaussian signaling system at low values of ${\bar \upsilon _{{{\mathrm{p}}}}}$. Although the proper Gaussian signaling system cannot get benefits from increasing the power budget, the improper Gaussian signaling tends to use the total budget efficiently and relieve the interference effect on PU by increasing $\mathcal{C}_x$, which compensates for the interference impact as can be seen in \eqref{pu_rate}. On the other hand, at high RSI-CNR values, both proper and improper fail to operate properly.                           

\begin{figure}[!t]
\centering
\includegraphics[width=9cm]{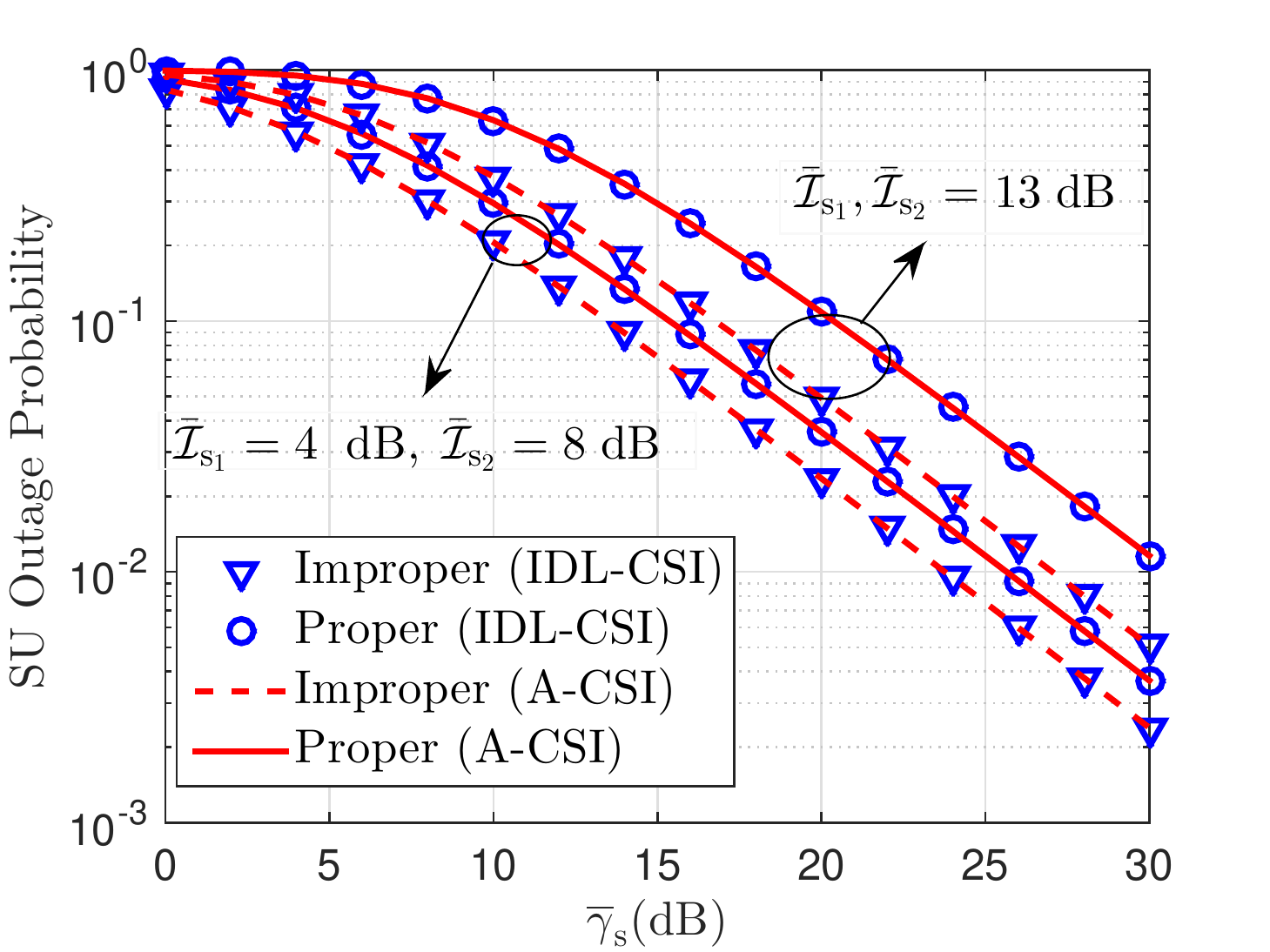}
\caption{SU outage probability for A-CSI and IDL-CSI based designs versus ${\bar \gamma _{{{\mathrm{s}}}}}$ for $\left(\bar {{\cal I}}_{{{\rm{s}}_1}},\bar {{\cal I}}_{{{\rm{s}}_2}}\right)=\left(4,8\right),\left(13,13\right)\;\rm{dB}$ pairs..}
\label{SimEx5}
\end{figure}

\textit{Example 5:} In this example, we compare the design of the proper/improper Gaussian signaling design based on both A-CSI and IDL-CSI. To this end, we plot SU outage probability versus ${\bar \gamma _{{{\mathrm{s}}}}}$ for different $\bar {{\cal I}}_{{{\rm{s}}_i}}$ in Fig. \ref{SimEx5}. The simulation result shows a perfect match between the A-CSI and IDL-CSI based designs for both the proper and improper Gaussian signaling schemes. For low $\bar{\gamma_{\mathrm{s}}}$, the SU outage occurs mainly because of its direct link and the IDL-CSI based design saves the SU power and stay silent. As a result, no data is delivered to the SU receiver and outage is reported. On the other hand, for large $\bar{\gamma_{\mathrm{s}}}$, the SU outage is mainly controlled by the PU interference link and the optimal signal parameters of IDL-CSI based system does not depend on $\bar{\gamma_{\mathrm{s}}}$ as can seen from  \eqref{opt_prob_equivalent}.

\begin{figure}[!t]
\centering
\includegraphics[width=9cm]{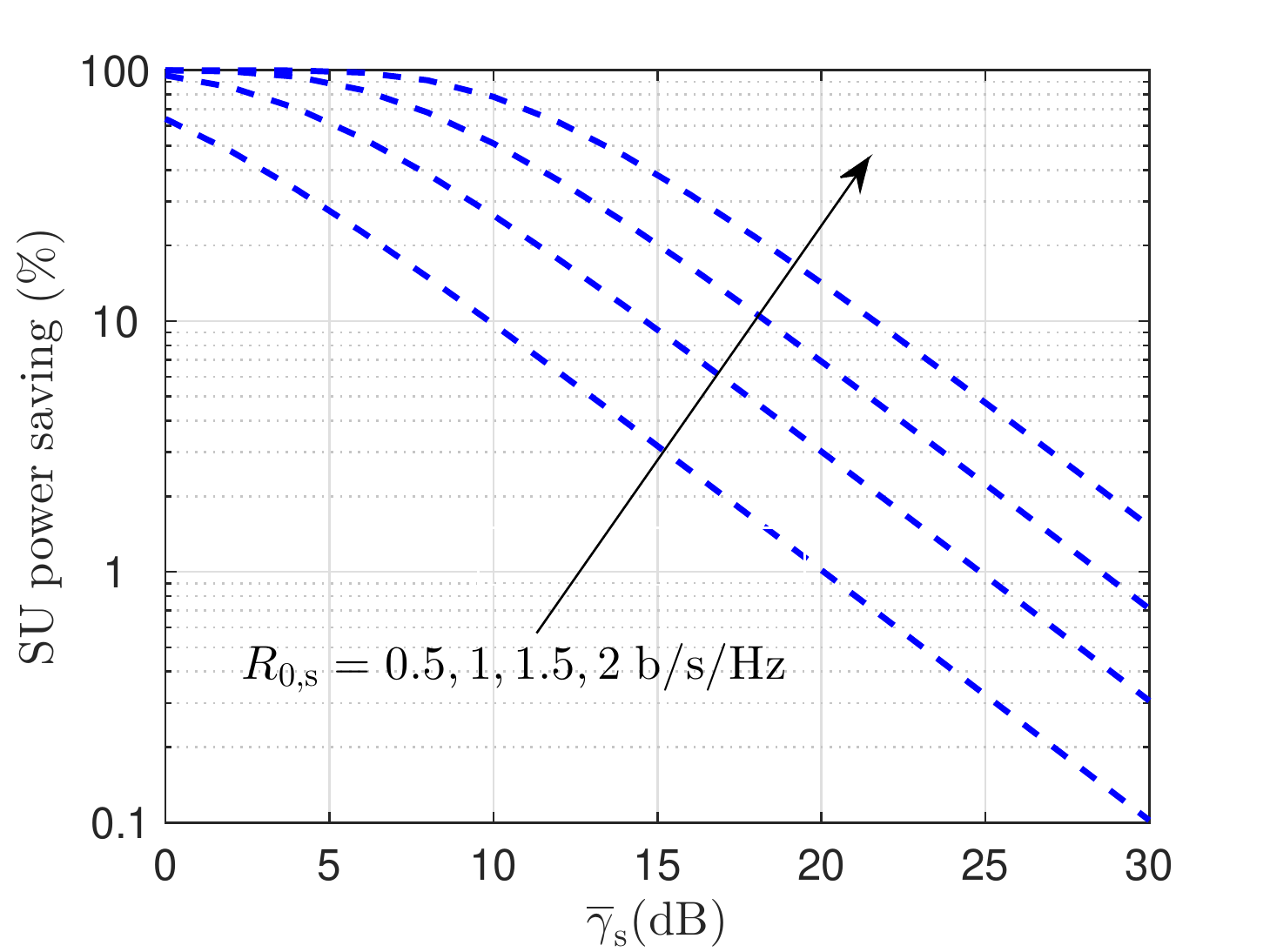}
\caption{Percentage of SU power saving of IDL-CSI improper based design versus ${\bar \gamma _{{{\mathrm{s}}}}}$ for $R_{\rm{0,s}}=0.5,1,1.5,2\;\rm{b/s/Hz}$.}
\label{SimEx6}
\end{figure}

\textit{Example 6:} This example studies the first benefit of knowledge of $\gamma_{\mathrm{s}}$ represented in the power saving. Fig. \ref{SimEx6} shows the SU power saving percentage versus ${\bar \gamma _{{{\mathrm{s}}}}}$ for different SU target rates of $R_{\rm{0,s}}$. It can be seen that at lower values of ${\bar \gamma _{{{\mathrm{s}}}}}$, the SU can save more power instead of transmitting and an outage event occurs. At higher values of ${\bar \gamma _{{{\mathrm{s}}}}}$, the SU channel conditions are good enough to achieve its target rate. Moreover, if the SU target rate increases, the probability of having an outage event increases and hence, the SU preferably stays idle and saves its transmit~power.  

\begin{figure}[!t]
\centering
\includegraphics[width=9cm]{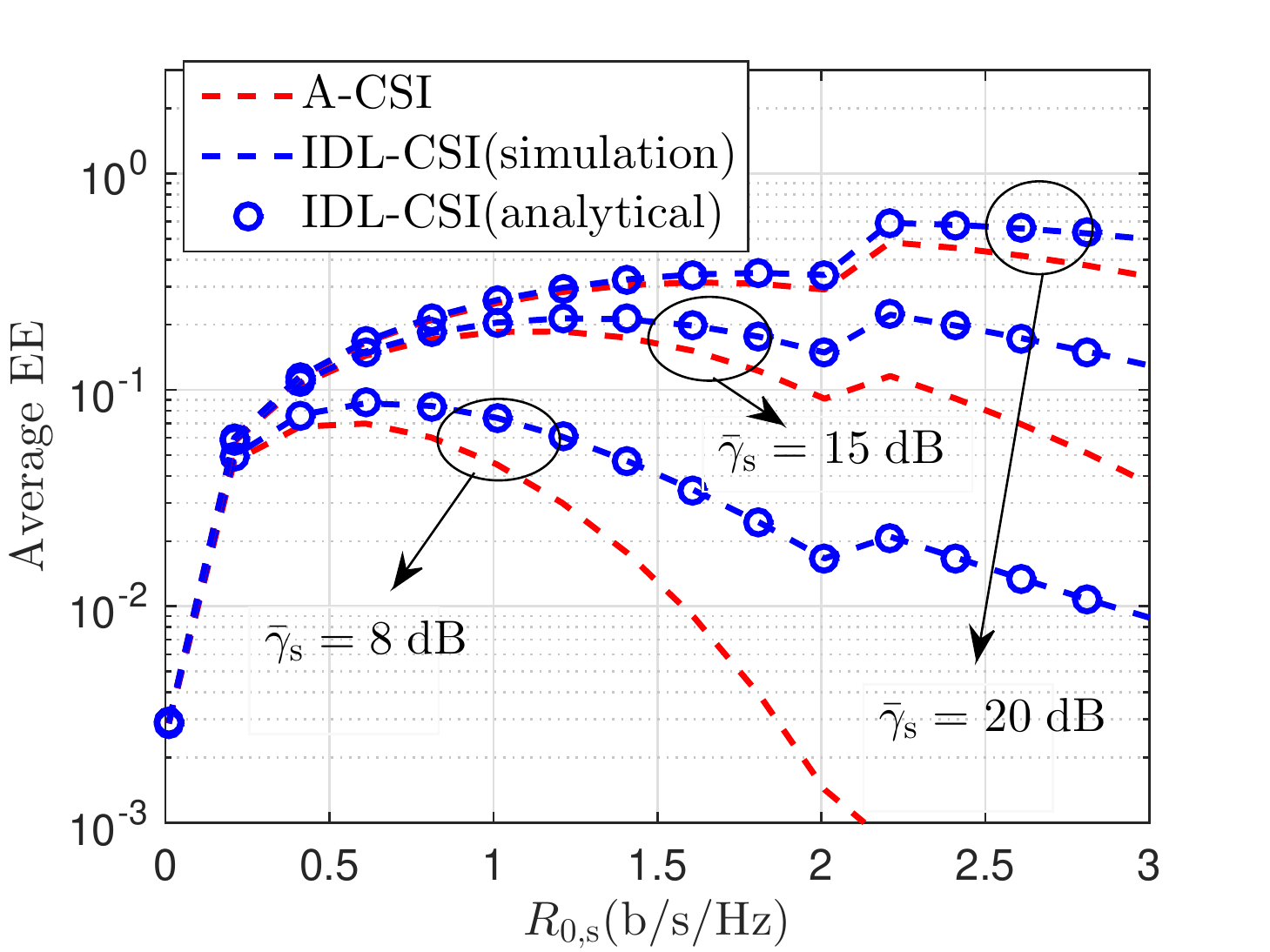}
\caption{SU average energy efficiency for A-CSI and IDL-CSI improper based designs versus $R_{\rm{0,s}}$ for ${\bar \gamma _{{{\mathrm{s}}}}}=8,15,20\;\rm{dB}$.}
\label{SimEx7}
\end{figure}

\textit{Example 7:} To investigate the average EE improvement benefit as a result of perfect knowledge of $\gamma_{\mathrm{s}}$, Fig. \ref{SimEx7} plots the SU average EE for the SU improper Gaussian signaling design based A-CSI and IDL-CSI using simulations and derived analytical expression \eqref{EE_avg_SU_expression}. We assume $\kappa = 5$ and $p_{\rm{c}}=1\;\rm{W}$. First, we observe that the simulation curve has a perfect match with the analytical expression. At lower SU target rates, the requirements for the SU are flexible and hence, both designs use less power and thus the average EE improves with increasing $\bar\gamma_{\mathrm{s}}$. On the other hand, as the SU rate increases, the requirements become more stringent, which force the SU to increase its transmit power and therefore, deteriorating the average EE performance. 
As $R_{\rm{0,s}}$ increases, the  IDL-CSI based design uses less power to achieve the same target rate. The gap between the two designs increases as ${\bar \gamma _{{{\mathrm{s}}}}}$ decreases and hence, the IDL-CSI based design can make use of the perfect knowledge of the CSI in order to save the transmit power and therefore boost the average EE of the SU system. We observe also from Fig. \ref{SimEx7} that as $R_{\rm{0,s}}$ increases, it reaches specific values at which the average EE behavior shows abrupt improvement. As we know from the improper condition \eqref{improper_condition} that there is a specific value of $R_{\rm{0,s}}$, at which improper signaling can not be used, thus the solution switches to proper signaling at larger values $R_{\rm{0,s}}$. Since the proper signaling scheme uses less power, it achieves better average EE which interprets this sudden improvement of the average EE performance. 

\begin{figure}[!t]
\centering
\includegraphics[width=9cm]{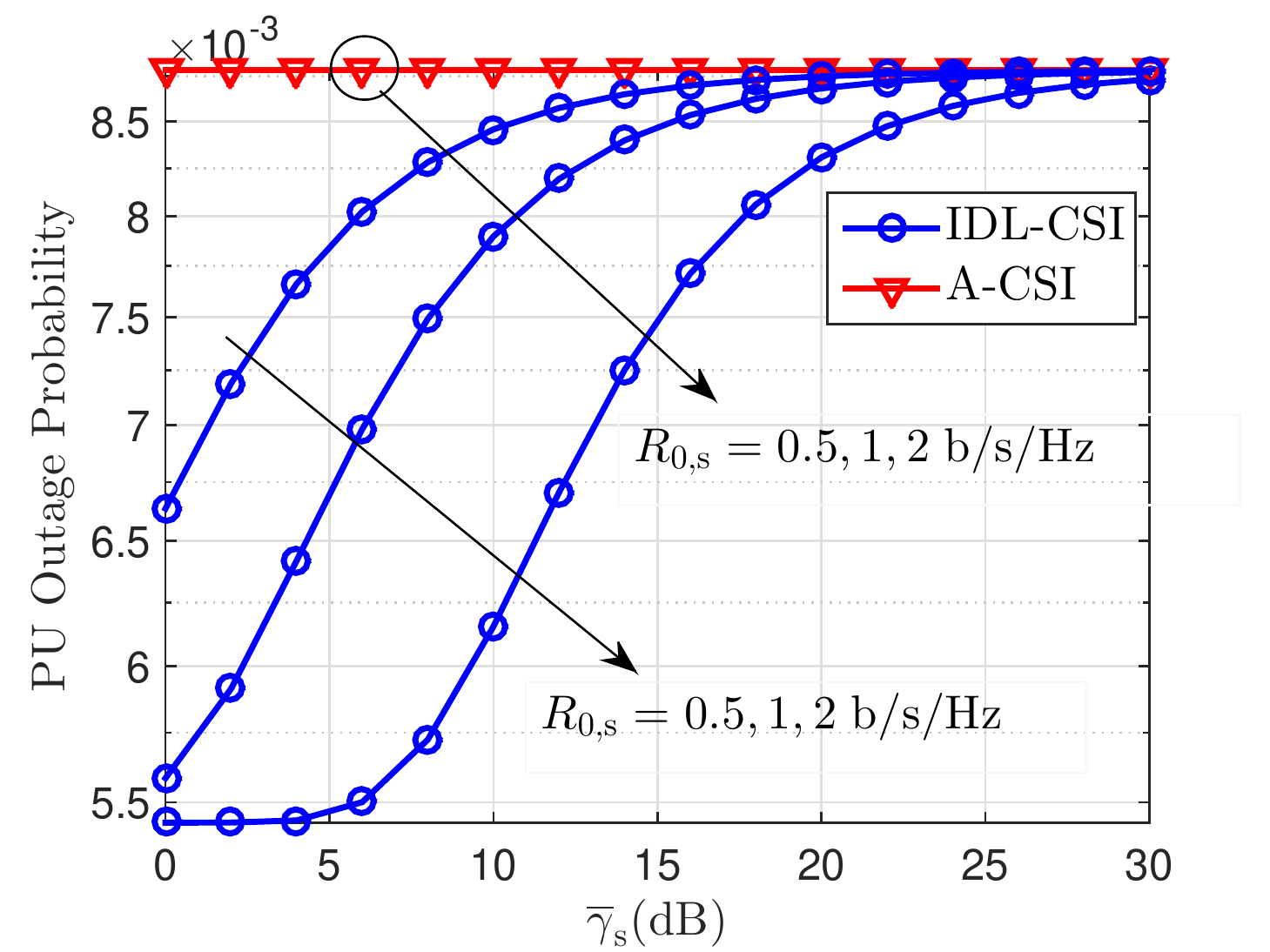}
\caption{PU outage probability for A-CSI and IDL-CSI based designs versus ${\bar \gamma _{{{\mathrm{s}}}}}$ for $R_{\rm{0,s}}=0.5,1,2\;\rm{b/s/Hz}$.}
\label{SimEx8}
\end{figure}

\textit{Example 8:} This example illustrates the benefits that can be reaped at the PU side by the design based on perfect knowledge of $\gamma_\mathrm{s}$. For this purpose, we plot the PU outage probability A-CSI and IDL-CSI based designs versus ${\bar \gamma _{{{\mathrm{s}}}}}$ for different $R_{\rm{0,s}}$ in Fig. \ref{SimEx8}. The A-CSI based design PU outage probability is fixed and does not change with ${\bar \gamma _{{{\mathrm{s}}}}}$ or $R_{\rm{0,s}}$ as expected. On the other hand, the IDL-CSI based design has different performance. Specifically, at low ${\bar \gamma _{{{\mathrm{s}}}}}$, $P_{\mathrm{saving}}$ has more value than higher  values of ${\bar \gamma _{{{\mathrm{s}}}}}$ as can be shown in Fig. \ref{SimEx6}. Thus, PU outage probability is expected to get more benefit as ${\bar \gamma _{{{\mathrm{s}}}}}$ decreases. 
Furthermore, As the SU target rate increases, the requirements for the SU become rigid and and the chances for the SU to stay idle become higher, which improves ~the PU outage probability. 
%
%

\section{Conclusion}
In this paper, we investigate the opportunity of sharing the spectrum resources of licensed FD PU in underlay CR mode using improper Gaussian signaling. We use  the outage probability as a performance metric, then  derive a closed form for the SU and a tight upper bound for the PU. Based on the average CSI, we optimize the SU signal parameters, i.e., transmit power and circularity coefficient, to minimize the SU outage probability while maintaining a predetermined QoS requirements for the PU. As a result, we derive a low complexity algorithm that tunes the signal parameters to accomplish the design objectives. Moreover, we show that improper Gaussian signaling is beneficial if the maximum allowable INR at the PU exceeds a predefined threshold and the SU works under a maximum allowable target rate. Then, we study the benefits of perfect knowledge of the SU direct link. Specifically, the SU can save more power and improve its average EE, and the PU can achieve more protection in terms of its outage probability performance. The later merit may  be used by other cognitive users in the same system to access the spectrum without deteriorating the PU QoS. The numerical results show a promising performance for the improper Gaussian signaling with A-CSI and IDL-CSI. Specifically, the main advantage of the proposed scheme is for strong SU interference channels to PU, where proper signaling scheme tends to use less power, while SU with improper Gaussian signaling uses more power and compensates its interference impact through the increase of signal ~impropriety. 
\section*{Appendix A}
In this appendix, we prove that {\footnotesize{$p_\mathrm{s}^{\left( i \right)}\left({\cal C}_x\right)$}} in \eqref{ps_improper_max} is strictly increasing in ${\cal C}_x$ over the interested interval {\footnotesize{$0 < {{\cal C}_x} < 1$}}. We assume here, that the SU is allowed to transmit, i.e., {\footnotesize{$\Upsilon _i>0$}}. The first derivative of {\footnotesize{$p_\mathrm{s}^{\left( i \right)}\left({\cal C}_x\right)$}} can be written as
\footnotesize
\begin{align}\label{ps_inc_proof}
\frac{{dp_{\rm{s}}^{\left( i \right)}\left( {{\cal C}{_x}} \right)}}{{d{{\cal C}_x}}} = \frac{{{{\cal C}_x}\left( {{\Phi _i} - 2{\Lambda _i}\left( { - {\Lambda _i} + \sqrt {\Lambda _i^2 + {\Phi _i}} } \right)} \right)}}{{{\Gamma _{{\mathrm{p}}_i}}{{\bar {\cal I}}_{{{\rm{s}}_j}}}{{\left( {1 - {\cal C}_x^2} \right)}^2}\sqrt {\Lambda _i^2 + {\Phi _i}} }},
\end{align}
\normalsize
where {\footnotesize{${\Phi _i} = {\Gamma _{{\mathrm{p}}_i}}\left( {1 - {\cal C}_x^2} \right){\Upsilon _i}$}}. Assuming that {\footnotesize{$R_{0,\rm{p}_i}>0$}}, hence {\footnotesize{${\Phi _i} > 0$}}. From \eqref{ps_inc_proof}, if {\footnotesize{${\Lambda _i}\leq 0$}}, it is always positive. On the other hand, if {\footnotesize{${\Lambda _i}>0$}} we can rewrite \eqref{ps_inc_proof} as
\footnotesize
\begin{align}\label{dervative_appendix_A}
\frac{{dp_{\rm{s}}^{\left( i \right)}\left( {{\cal C}{_x}} \right)}}{{d{{\cal C}_x}}} = \frac{{{{\cal C}_x}{\Lambda _i}\left( {\left( {{\Omega _i} + 2} \right) - \sqrt {{{\left( {{\Omega _i} + 2} \right)}^2} - \Omega _i^2} } \right)}}{{{\Gamma _{{\mathrm{p}}_i}}{{\bar {\cal I}}_{{{\rm{s}}_j}}}{{\left( {1 - {\cal C}_x^2} \right)}^2}\sqrt {1 + {\Omega _i}} }},
\end{align}
\normalsize
where {\footnotesize{${\Omega _i} = {\Phi _i}/\Lambda _i^2$}}. It is clear that \eqref{dervative_appendix_A} is always positive and this concludes the proof.     
\section*{Appendix B}
In this appendix, we derive the conditions in \eqref{improper_condition} over the interested interval {\footnotesize{$0 < {{\cal C}_x} < 1$}} by using the chain rule of partial derivatives for {\footnotesize{${P_{{\rm{out,s}}}}\left( {\mathcal{G}_m,\mathcal{Y}} \right)$, $j \neq m$}}, in \eqref{p_out_s_Cx} with respect to {\footnotesize{${{\cal C}_x}$}},
\footnotesize
\begin{align}
\frac{{\partial {P_{{\rm{out,s}}}}\left( {\mathcal{G}_m,\mathcal{Y}} \right)}}{{\partial {{\cal C}_x}}} = \frac{{\partial {P_{{\rm{out,s}}}}}}{{\partial \mathcal{G}_m}}\frac{{d\mathcal{G}_m}}{{d{{\cal C}_x}}} + \frac{{\partial {P_{{\rm{out,s}}}}}}{{\partial \mathcal{Y}}}\frac{{d\mathcal{Y}}}{{d{{\cal C}_x}}},
\end{align} 
\normalsize
which is simplified to be
\footnotesize
\begin{align}\label{der_pout_c}
&\frac{{\partial {P_{{\rm{out,s}}}}\left( {\mathcal{G}_m,\mathcal{Y}} \right)}}{{\partial {{\cal C}_x}}} = \left( {\underbrace {\frac{{{ T}{{\cal C}_x}{{\bar \gamma }_{\rm{s}}}\left( { - {\Lambda _m} + \sqrt {\Lambda _m^2 + {\Phi _m}} } \right)}}{{{\Gamma _{{p_j}}}{{\bar {\cal I}}_{{{\rm{s}}_j}}}\left( {\sqrt {1 + \left( {1 - {\cal C}_x^2} \right){\Gamma _{\rm{s}}}}  - 1} \right)\left( {1 - {\cal C}_x^2} \right)}}}_{ \geq 0}} \right) \left( {\underbrace {\frac{{{\Lambda _m}}}{{\sqrt {\Lambda _m^2 + {\Phi _m}} }} - \frac{1}{{\sqrt {1 + \left( {1 - {\cal C}_x^2} \right){\Gamma _{\rm{s}}}} }}}_{a}} \right),
\end{align}
\normalsize
where
\footnotesize
\begin{align}
T = &\frac{{\mathcal{Y}\mathcal{G}_m\left(\bar\gamma_{\rm{s}}\right)\left( {\prod\limits_{j = 1}^2 {{p_j}{{\bar {\cal I}}_{{{{\rm{p}}}_j}}}}  + \Theta  - {\mathcal{Y}^2{\mathcal{G}_m^2\left(\bar\gamma_{\rm{s}}\right)}}} \right) + \Theta }}{{{\Theta ^2}}} \exp \left( { - \frac{1}{\mathcal{Y}\mathcal{G}_m\left(\bar\gamma_{\rm{s}}\right)}} \right).
\end{align}
\normalsize
and {\footnotesize{$\Theta  = \prod\limits_{j = 1}^2 {\left( {{p_j}{{\bar {\cal I}}_{{{{\rm{p}}}_j}}} + \mathcal{Y}\mathcal{G}_m\left(\bar\gamma_{\rm{s}}\right)} \right)}$}}. Here, we have two cases for {\footnotesize{$\Lambda_m$}}. If {\footnotesize{$\Lambda_m \leq 0$}}, then {\footnotesize{$a \leq 0$}} and \eqref{der_pout_c} is non-positive, hence, {\footnotesize{${P_{{\rm{out,s}}}}\left(\mathcal{C}_x\right)$}} is monotonically decreasing in {\footnotesize{$\mathcal{C}_x$}}. On the other hand, if {\footnotesize{$\Lambda_m > 0 $}}, then one can deduce easily that when {\footnotesize{${\Lambda _m} \leq \sqrt {{\Gamma _{{{\rm{p}}_m}}}{\Upsilon _m}/{\Gamma _{\rm{s}}}} $}},  {\footnotesize{$a \leq 0$}} and hence, {\footnotesize{${P_{{\rm{out,s}}}}\left(\mathcal{C}_x\right)$}} is monotonically decreasing in {\footnotesize{$\mathcal{C}_x$}}. Otherwise, if {\footnotesize{${\Lambda _m} > \sqrt {{\Gamma _{{{\rm{p}}_m}}}{\Upsilon _m}/{\Gamma _{\rm{s}}}} $}}, {\footnotesize{${P_{{\rm{out,s}}}}\left(\mathcal{C}_x\right)$}} is  monotonically increasing in {\footnotesize{$\mathcal{C}_x$}}. It is clear that the condition {\footnotesize{${\Lambda _m} \leq \sqrt {{\Gamma _{{{\rm{p}}_m}}}{\Upsilon _m}/{\Gamma _{\rm{s}}}} $}} combines both cases of {\footnotesize{$\Lambda_m$}} for {\footnotesize{${P_{{\rm{out,s}}}}\left(\mathcal{C}_x\right)$}} to be monotonically decreasing in {\footnotesize{$\mathcal{C}_x$}}. Furthermore, this condition can be rewritten as
\footnotesize    
\begin{equation}\label{R0_s_condition}
{R_{0,{\rm{s}}}} \leq \frac{1}{2}{\log _2}\left( {1 + \frac{{{\Gamma _{{\rm{p}_m}}}{\Upsilon _m}}}{{\Lambda _m^2}}} \right),
\end{equation} 
\normalsize 
which can be further simplified by inserting the expressions for {\footnotesize{${\Upsilon _m}$, ${\Lambda _m}$}} giving the condition \eqref{improper_condition} and concludes the proof.
\section*{Appendix C}
In this appendix, we prove the monotonicity conditions of {\footnotesize{$P_{{\rm{out,s}}}^{\mathrm{DL-T}}\left( {{{\cal G}_m},{\cal Y}} \right)$}}, {\footnotesize{$j \neq m$}}, in \eqref{p_out_s_DL_Cx} over the interested interval {\footnotesize{$0 < {{\cal C}_x} < 1$}}. We assume here that the transmission condition in \eqref{trans_cond} is satisfied, i.e., the SU is allowed to transmit. By using the chain rule of partial derivatives as in Appendix B, we get
\footnotesize
\begin{align}\label{dervative_pouts_DL}
\frac{\partial P_{{\rm{out,s}}}^{\mathrm{DL-T}}\left( {{{\cal G}_m},{\cal Y}} \right)}{\partial{\cal C}{_x}} &=M \times \underbrace{\frac{{{\cal C}{_x}{\gamma _{\rm{s}}}\left( { - {\Lambda _m} + \sqrt {\Lambda _m^2 + {\Phi _m}} } \right)}}{{{\Gamma _{{p_m}}}{{\bar {\cal I}}_{{{\rm{s}}_j}}}\left( {\sqrt {1 + \left( {1 - {\cal C}_x^2} \right){\Gamma _{\rm{s}}}}  - 1} \right)\left( {1 - {\cal C}_x^2} \right)}}}_{\geq 0}   \left( \underbrace{ {\frac{{{\Lambda _m}}}{{\sqrt {\Lambda _m^2 + {\Phi _m}} }} - \frac{1}{{\sqrt {1 + \left( {1 - {\cal C}_x^2} \right){\Gamma _{\rm{s}}}} }}}}_{a} \right),
\end{align} \normalsize   
where
\footnotesize
\begin{equation}\label{M}
M = \left\{ {\begin{array}{*{20}{c}}
\sum\limits_{i = 1\atop
j \ne i}^2 {\frac{{{\cal K}{_{i,j}}}}{{{p_i}{{\bar {\cal I}}_{{{\rm{p}}_i}}}}}\exp \left( { - \frac{ {{\cal Y}{{\cal G}_m}\left( {{\gamma _{\rm{s}}}} \right) - 1} }{{{p_i}{{\bar {\cal I}}_{{{\rm{p}}_i}}}}}} \right)},&\rm{if}\;\;{{{p_1}{{\bar {\cal I}}_{{{{\rm{p}}}_1}}} \neq {p_2}{{\bar {\cal I}}_{{{{\rm{p}}}_2}}}}} \\  \\
\left( {\frac{{{\cal Y}{{\cal G}_m}\left( {{\gamma _{\rm{s}}}} \right) - 1}}{{{{\left( {{p_1}{{\bar {\cal I}}_{{{\rm{p}}_1}}}} \right)}^2}}}} \right)\exp \left( { - \frac{{{\cal Y}{{\cal G}_m}\left( {{\gamma _{\rm{s}}}} \right) - 1}}{{{p_1}{{\bar {\cal I}}_{{{\rm{p}}_1}}}}}} \right),&\rm{if}\;\;{{{p_1}{{\bar {\cal I}}_{{{{\rm{p}}}_1}}} = {p_2}{{\bar {\cal I}}_{{{{\rm{p}}}_2}}}}}
\end{array}} \right..
\end{equation}
\normalsize 
For instance, if {\footnotesize{$M$}} is non-negative  and when the condition in \eqref{improper_condition} is valid, then, {\footnotesize{$a \leq 0$}}. As a result,   \eqref{dervative_pouts_DL} is non-positive and {\footnotesize{$P_{{\rm{out,s}}}^{\mathrm{DL-T}}$}} is monotonically decreasing in {\footnotesize{$\mathcal{C}_x$}}. On the other hand, if the condition in \eqref{improper_condition} does not hold,  then {\footnotesize{$a > 0$}} and therefore, {\footnotesize{$P_{{\rm{out,s}}}^{\mathrm{DL-T}}$}} is monotonically increasing in {\footnotesize{$\mathcal{C}_x$}}. To this point, we need now to show that {\footnotesize{$M$}} is non-negative. For asymmetric PU interference links, \eqref{M} can be expressed as
\footnotesize
\begin{align}\label{M_Appendix}
M = \frac{1}{{{p_1}{{\bar {\cal I}}_{{{\rm{p}}_1}}} - {p_2}{{\bar {\cal I}}_{{{\rm{p}}_2}}}}} \times \Bigg[& \exp \left( { - \frac{{{\cal Y}{{\cal G}_m}\left( {{\gamma _{\rm{s}}}} \right) - 1}}{{{p_1}{{\bar {\cal I}}_{{{\rm{p}}_1}}}}}  } \right) -   \exp \left( { - \frac{{{\cal Y}{{\cal G}_m}\left( {{\gamma _{\rm{s}}}} \right) - 1}}{{{p_2}{{\bar {\cal I}}_{{{\rm{p}}_2}}}}}  } \right) \Bigg].
\end{align}
\normalsize  
Here, we have two cases, either {\footnotesize{${p_1}{{\bar {\cal I}}_{{{\rm{p}}_1}}} > {p_2}{{\bar {\cal I}}_{{{\rm{p}}_2}}}$}} or {\footnotesize{${p_1}{{\bar {\cal I}}_{{{\rm{p}}_1}}} < {p_2}{{\bar {\cal I}}_{{{\rm{p}}_2}}}$}}. For the first,  it is clear that {\footnotesize{${\rm M} \geq 0$}}. For the second, the same argument applies by replacing every index $1$ by $2$ and vice versa in \eqref{M_Appendix}. For the other scenario, when {\footnotesize{${{{p_1}{{\bar {\cal I}}_{{{{\rm{p}}}_1}}} = {p_2}{{\bar {\cal I}}_{{{{\rm{p}}}_2}}}}}$}}, we can write {\footnotesize{$M$}} as
\footnotesize
\begin{align}
M = \frac{{\zeta \left( {p_{\rm{s}}^{\left( m \right)},{{\cal C}_x}} \right)}}{{{{\left( {{p_1}{{\bar {\cal I}}_{{{\rm{p}}_1}}}} \right)}^2}}}\exp \left( { - \frac{{\zeta \left( {p_{\rm{s}}^{\left( m \right)},{{\cal C}_x}} \right)}}{{{{\left( {{p_1}{{\bar {\cal I}}_{{{\rm{p}}_1}}}} \right)}^2}}}} \right),
\end{align}  
\normalsize  
which is found to be non-negative by following  \eqref{trans_cond} and this concludes the proof.

\section*{Appendix D}
In this appendix, we aim at deriving the equivalent optimization problem in \eqref{opt_prob_equivalent}. We assume here that the transmission condition in \eqref{trans_cond} is fulfilled, i.e., the SU is allowed to transmit. First, we can write {\footnotesize{$ P_{{\rm{out,s}}}^{{\rm{DL-T}}}\left( {p_o^{(z)},{\cal C}_o^{(z)}} \right)$}} in \eqref{opt_prob_global_inst}, when {\footnotesize{${{{p_1}{{\bar {\cal I}}_{{{{\rm{p}}}_1}}} \neq {p_2}{{\bar {\cal I}}_{{{{\rm{p}}}_2}}}}}$}}, as
\footnotesize
\begin{align}\label{Appendix_p_out_inst}
\!\!\!\!\!\!\!\! P_{{\rm{out,s}}}^{{\rm{DL-T}}}\left( {p_o^{(z)},{\cal C}_o^{(z)}} \right) = \frac{1}{{{p_1}{{\bar {\cal I}}_{{{\rm{p}}_1}}} - {p_2}{{\bar {\cal I}}_{{{\rm{p}}_2}}}}}    \Bigg[ {{p_1}{{\bar {\cal I}}_{{{\rm{p}}_1}}}\exp \left( { - \frac{{{\gamma _{\rm{s}}}\Delta  - 1}}{{{p_1}{{\bar {\cal I}}_{{{\rm{p}}_1}}}}}  } \right)} - {p_2}{{\bar {\cal I}}_{{{\rm{p}}_2}}}\exp \Big( - \frac{{{\gamma _{\rm{s}}}\Delta  - 1}}{{{p_2}{{\bar {\cal I}}_{{{\rm{p}}_2}}}}}\Big) \Bigg],
\end{align}
\normalsize
where {\footnotesize{$\Delta  = \frac{{\left( {1 - {{\left( {{\cal C}_o^{(z)}} \right)}^2}} \right)}}{{{\Psi _{\rm{s}}}\left( {p_o^{(z)},{\cal C}_o^{(z)}} \right)}}$}}. Similar to the proof in Appendix C, we assume, first, that {\footnotesize{${p_1}{{\bar {\cal I}}_{{{\rm{p}}_1}}} > {p_2}{{\bar {\cal I}}_{{{\rm{p}}_2}}}$}}. We calculate the first derivative of \eqref{Appendix_p_out_inst} with respect to {\footnotesize{$\Delta$}} as  
\footnotesize   
\begin{align}\label{Appendix_p_out_inst_dervative}
&\frac{{dP_{{\rm{out,s}}}^{{\rm{DL-T}}}}}{{d\Delta }} = \frac{{{\gamma _{\rm{s}}}}}{{{p_1}{{\bar {\cal I}}_{{{\rm{p}}_1}}} - {p_2}{{\bar {\cal I}}_{{{\rm{p}}_2}}}}}\times \Bigg[ \exp \left( { - \frac{{{\gamma _{\rm{s}}}\Delta  - 1}}{{{p_2}{{\bar {\cal I}}_{{{\rm{p}}_2}}}}}  } \right) - \exp \left( { - \frac{{{\gamma _{\rm{s}}}\Delta  - 1}}{{{p_1}{{\bar {\cal I}}_{{{\rm{p}}_1}}}}}  } \right) \Bigg],
\end{align}
\normalsize 
which is always non-positive, independent of the value of {\footnotesize{${{\gamma _{\rm{s}}}}$}}, and hence, {\footnotesize{$ P_{{\rm{out,s}}}^{{\rm{DL-T}}}\left( {p_o^{(z)},{\cal C}_o^{(z)}} \right)$}} is monotonically decreasing in {\footnotesize{$\Delta$}}. Thus, we can replace the original optimization problem by maximizing {\footnotesize{$\Delta$}} over the optimal local solution pairs {\footnotesize{$\left( {p_o^{(z)},{\cal C}_o^{(z)}} \right)$}} to obtain the global solution pair {\footnotesize{$\left( {p_{\rm{s}}^*,{\cal C}_x^*} \right)$}} which yields \eqref{opt_prob_equivalent}. Also, For the other case in which  {\footnotesize{${p_1}{{\bar {\cal I}}_{{{\rm{p}}_1}}} < {p_2}{{\bar {\cal I}}_{{{\rm{p}}_2}}}$}}, the same argument applies as illustrated in Appendix C. On the other hand, when {\footnotesize{${{{p_1}{{\bar {\cal I}}_{{{{\rm{p}}}_1}}} = {p_2}{{\bar {\cal I}}_{{{{\rm{p}}}_2}}}}}$}}, we can write the derivative~as
\footnotesize
\begin{align}
\frac{{dP_{{\rm{out}},{\rm{s}}}^{{\rm{DL - T}}}}}{{d\Delta }} =  - \frac{{{\gamma _{\rm{s}}}}}{{{p_1}{{\bar {\cal I}}_{{{\rm{p}}_1}}}}}\left( {\frac{{{\gamma _{\rm{s}}}\Delta  - 1}}{{{p_1}{{\bar {\cal I}}_{{{\rm{p}}_1}}}}}} \right)\exp \left( { - \frac{{{\gamma _{\rm{s}}}\Delta  - 1}}{{{p_1}{{\bar {\cal I}}_{{{\rm{p}}_1}}}}}} \right)
\end{align}
\normalsize 
which, from the transmission condition in \eqref{trans_cond}, it follows that the derivative is non-positive hence, same illustrated arguments apply and this concludes the proof.

\bibliographystyle{IEEEtran}

\bibliography{IEEEabrv,mgaafar_ref_October_2015}

\end{document}